\documentclass[journal]{IEEEtran}
\usepackage[T1]{fontenc}
\usepackage{blindtext}
\usepackage[utf8]{inputenc}
\usepackage[noadjust]{cite}
\usepackage[font=footnotesize]{caption}
\usepackage{dblfloatfix}
\usepackage{flafter}
\usepackage{graphicx}
\usepackage{psfrag}
\usepackage{subfig}
\usepackage{amsmath,amssymb,amsthm,mathrsfs,amsfonts,dsfont}
\usepackage{color}
\usepackage[algoruled]{algorithm2e}
\usepackage{fixltx2e}
\usepackage{multirow}
\usepackage{multicol}
\usepackage[hidelinks]{hyperref}
\usepackage[normalem]{ulem}
\usepackage{acronym}
\usepackage[table,xcdraw]{xcolor}
\usepackage{accents}
\usepackage{mwe}
\usepackage{hhline}
\usepackage{trfsigns}
\usepackage{siunitx}
\usepackage{lscape}
\usepackage{wasysym}
\usepackage{bm}

\newacro{3GPP}{3rd Generation Partnership Project}
\newacro{5G}{fifth generation}
\newacro{6G}{sixth generation}
\newacro{A/D}{analog-to-digital}
\newacro{ABE}{analog back-end}
\newacro{ADC}{analog-to-digital converter}
\newacro{AFE}{analog front-end}
\newacro{AGV}{automatic guided vehicle}
\newacro{AWGN}{additive white Gaussian noise}
\newacro{B5G}{beyond \ac{5G}}
\newacro{BB}{baseband}
\newacro{BER}{bit error ratio}
\newacro{BPSK}{binary phase-shift keying}
\newacro{BP}{band-pass}
\newacro{BS}{base station}
\newacro{CDM}{code-division multiplexing}
\newacro{CFO}{carrier frequency offset}
\newacro{CFR}{channel frequency response}
\newacro{CIR}{channel impulse response}
\newacro{CoMP}{coordinated multipoint}
\newacro{CP}{cyclic prefix}
\newacro{CPE}{common phase error}
\newacro{CPO}{carrier phase offset}
\newacro{CRLB}{Cram\'er?Rao lower bound}
\newacro{CS}{chirp sequence}
\newacro{CW}{continuous wave}
\newacro{CZT}{chirp Z-transform}
\newacro{D/A}{digital-to-analog}
\newacro{DAC}{digital-to-analog converter}
\newacro{DDS}{direct digital synthesis}
\newacro{DFRC}{dual-function radar-cunication or dual-functional radar-cunication}
\newacro{DFnT}{discrete Fresnel transform}
\newacro{DFT}{discrete Fourier transform}
\newacro{DMRS}{demodulation reference signal}
\newacro{DoA}{direction-of-arrival}
\newacro{DoD}{direction-of-departure}
\newacro{DPD}{digital pre-distortion}
\newacro{ETSI}{European Telecommunications Standards Institute}
\newacro{EVM}{error vector magnitude}
\newacro{FDE}{frequency-domain equalization}
\newacro{FDM}{frequency-division multiplexing}
\newacro{FO}{frequency offset}
\newacro{gNB}{gNodeB}
\newacro{HP}{high-pass}
\newacro{IBFD}{in-band full duplex}
\newacro{ICI}{intercarrier interference}
\newacro{IDFT}{inverse discrete Fourier transform}
\newacro{IDFnT}{inverse discrete Fresnel transform}
\newacro{IF}{intermediate frequency}
\newacro{IHE}{Institute of Radio Frequency Engineering and Electronics}
\newacro{I/Q}{in-phase/quadrature}
\newacro{ISAC}{integrated sensing and communication}
\newacro{ISI}{intersymbol interference}
\newacro{ISLR}{integrated-sidelobe level ratio}
\newacro{JCAS}{joint communication and sensing}
\newacro{KIT}{Karlsruhe Institute of Technology}
\newacro{LDPC}{low-density parity-check}
\newacro{LFSR}{linear-feedback shift register}
\newacro{LNA}{low-noise amplifier}
\newacro{LO}{local oscillator}
\newacro{LoS}{line-of-sight}
\newacro{LP}{low-pass}
\newacro{LS}{least squares}
\newacro{mmWave}{milimeter wave}
\newacro{MIMO}{multiple-input multiple-output}
\newacro{MLE}{maximum likelihood estimator}
\newacro{MLS}{maximum-length sequence}
\newacro{MRC}{maximal-ratio combining}
\newacro{MUSIC}{multiple signal classification}
\newacro{NLoS}{non-line-of-sight}
\newacro{NR}{new radio}
\newacro{OCDM}{orthogonal chirp-division multiplexing}
\newacro{OFDM}{orthogonal frequency-division multiplexing}
\newacro{OOB}{out-of-band}
\newacro{OTA}{over-the-air}
\newacro{P/S}{parallel-to-serial}
\newacro{PA}{power amplifier}
\newacro{PACF}{periodic autocorrelation function}
\newacro{PCCF}{periodic cross-correlation function}
\newacro{PLC}{powerline cunication}
\newacro{PLL}{phase-locked loop}
\newacro{PMCW}{phase-modulated continuous wave}
\newacro{PMN}{perceptive mobile network}
\newacro{PN}{oscillator phase noise}
\newacro{PoC}{proof-of-concept}
\newacro{PPLR}{peak power loss ratio}
\newacro{PRBS}{pseudorandom binary sequence}
\newacro{PRS}{positioning reference signal}
\newacro{PSLR}{peak-to-sidelobe level ratio}
\newacro{QPSK}{quadrature phase-shift keying}
\newacro{RadCom}{radar-cunication}
\newacro{RCS}{radar cross section}
\newacro{RF}{radio-frequency}
\newacro{RIS}{reflective intelligent surface}
\newacro{RMSE}{root mean squared error}
\newacro{SC}[S\&C]{Schmidl \& Cox}
\newacro{SFO}{sampling frequency offset}
\newacro{SIC}{self-interference cancellation}
\newacro{SINR}{signal-to-interference-plus-noise ratio}
\newacro{SIR}{signal-to-interference ratio}
\newacro{SISO}{single-input single-output}
\newacro{SNR}{signal-to-noise ratio}
\newacro{SoC}{system-on-a-chip}
\newacro{SSB}{synchronization signal block}
\newacro{STO}{symbol time offset}
\newacro{S/P}{serial-to-parallel}
\newacro{TDE}{time-domain equalization}
\newacro{TDM}{time-division multiplexing}
\newacro{TDR}{time-domain reflectometry}
\newacro{TITO}{tilt inference of time offset}
\newacro{TO}{time offset}
\newacro{UE}{user equipment}
\newacro{ULA}{uniform linear array}
\newacro{UWAC}{underwater acoustic cunication}
\newacro{V2V}{vehicle-to-vehicle}
\newacro{ZF}{zero forcing}
\newacro{ZP}{zero padding}

\makeatletter
\renewcommand*\env@cases[1][1.2]{%
	\let\@ifnextchar\new@ifnextchar
	\left\lbrace
	\def\arraystretch{#1}%
	\array{@{}l@{\quad}l@{}}%
}
\makeatother
\begin{document}
	
\title{System Concept and Demonstration of\\ Bistatic MIMO-OFDM-based ISAC}

\author{Lucas Giroto de Oliveira, Xueyun Long, 		Christian Karle, Umut Utku Erdem, Taewon Jeong,\\ Elizabeth Bekker, Yueheng Li, Thomas Zwick, and Benjamin Nuss
	\thanks{The authors acknowledge the financial support by the Federal Ministry of Education and Research of Germany in the projects ``KOMSENS-6G'' (grant number: 16KISK123) and ``Open6GHub'' (grant number: 16KISK010). \textit{(Corresponding author: Lucas Giroto de Oliveira.)}}
	\thanks{L. Giroto de Oliveira, {X. Long}, {U. U. Erdem}, {T. Jeong}, {E. Bekker}, {T. Zwick}, and {B. Nuss} are with the Institute of Radio Frequency Engineering and Electronics (IHE), Karlsruhe Institute of Technology (KIT), 76131 Karlsruhe, Germany (e-mail: \mbox{lucas.oliveira@kit.edu}, \mbox{xueyun.long@kit.edu}, \mbox{umut.erdem@kit.edu}, \mbox{taewon.jeong@kit.edu}, \mbox{elizabeth.bekker@kit.edu}, \mbox{thomas.zwick@kit.edu}, \mbox{benjamin.nuss@kit.edu}).}
	\thanks{C. Karle is with the Institute for Information Processing Technology (ITIV), Karlsruhe Institute of Technology (KIT), 76131 Karlsruhe, Germany (e-mail: \mbox{christian.karle@kit.edu}).}
	\thanks{Y. Li is was with the Institute of Radio Frequency Engineering and Electronics (IHE), Karlsruhe Institute of Technology (KIT), 76131 Karlsruhe, Germany. He is now with the Institute of Intelligent Communication Technology, Shandong University (SDU), 250100 Jinan, China (e-mail: \mbox{yueheng.li@sdu.edu.cn}).}
}


\maketitle

\begin{abstract}
	In future sixth-generation (6G) mobile networks, radar sensing is expected to be offered as an additional service to its original purpose of communication. Merging these two functions results in integrated sensing and communication (ISAC) systems. In this context, bistatic ISAC appears as a possibility to exploit the distributed nature of cellular networks while avoiding highly demanding hardware requirements such as full-duplex operation. Recent studies have introduced strategies to perform required synchronization and data exchange between nodes for bistatic ISAC operation, based on orthogonal frequency-division multiplexing (OFDM), however, only for single-input single-output architectures. In this article, a system concept for a bistatic multiple-input multiple-output (MIMO)-OFDM-based ISAC system with beamforming at both transmitter and receiver is proposed, and a distribution synchronization concept to ensure coherence among the different receive channels for direction-of-arrival estimation is presented. After a discussion on the ISAC processing chain, including relevant aspects for practical deployments such as transmitter digital pre-distortion and receiver calibration, a $4\times8$ MIMO measurement setup at $\SI{27.5}{\giga\hertz}$ and results are presented to validate the proposed system and distribution synchronization concepts.
\end{abstract}


\IEEEpeerreviewmaketitle


\section{Introduction}\label{sec:introduction}

\IEEEPARstart{A}{s} the development of \ac{6G} cellular networks progresses \cite{viswanathan2020}, \ac{ISAC} \cite{giroto2021_tmtt,liu2022} is emerging as a disruptive feature, with its inclusion anticipated in forthcoming \ac{3GPP} specifications. The introduction of \ac{ISAC} will extend the role of the network beyond communication, allowing it to operate as a pervasive radar sensing infrastructure \cite{prelcic2024,mandelli2023survey,kadelka2023,liu2022}. By integrating radar sensing into existing infrastructure, hardware and spectral resources can be shared while simultaneously allowing data communication and situational awareness in a wide range of use cases. These include, for example, object detection and tracking in indoor scenarios, environment monitoring in urban scenarios, and motion monitoring in healthcare or gesture recognition applications as recently discussed by \ac{ETSI} \cite{ETSIGRISC001} as well as by industry and academic entities \cite{kadelka2023,shatov2024}.

Aiming to exploit the inherently distributed nature of cellular networks for sensing \cite{su2024,thomae2023,ksiezyk2023} and avoid demanding hardware requirements such as \ac{IBFD} operation, recent studies in the literature have intensively investigated bistatic sensing both based on correlation of sensing paths with a reference path \cite{thomae2019,samczynski2022,elgamal2025} and using classical radar processing after estimating the transmit data \cite{giroto2023_EuMW,brunner2024} with the goal of ultimately enabling multistatic \ac{ISAC} networks \cite{mollen2023,su2024}. For that purpose, efforts are being concentrated not only on the development of channel models \cite{luo2024,andrich2024}, \ac{DoD} and \ac{DoA} estimation techniques \cite{naoumi2024}, and imaging techniques \cite{fenske2024}, but also on hardware impairments \cite{koivunen2024,giroto2024PN,mateosramos2025} and synchronization approaches \cite{pegoraro2024,aguilar2024,kai2024,han2025}.

\begin{figure*}[!t]
	\centering
	
	\psfrag{A}[c][c]{\footnotesize $R^\text{Tx---T}_p$}
	\psfrag{B}[c][c]{\footnotesize $R^\text{T---Rx}_p$}
	\psfrag{C}[c][c]{\footnotesize $R_0$}
	\psfrag{D}[c][c]{\footnotesize $x[s]$}
	\psfrag{E}[c][c]{\footnotesize $y[s]$}
	\psfrag{F}[c][c]{\footnotesize $F_\mathrm{s}^\mathrm{Tx}$}
	\psfrag{G}[c][c]{\footnotesize $F_\mathrm{s}^\mathrm{Rx}$}
	\includegraphics[height=9cm]{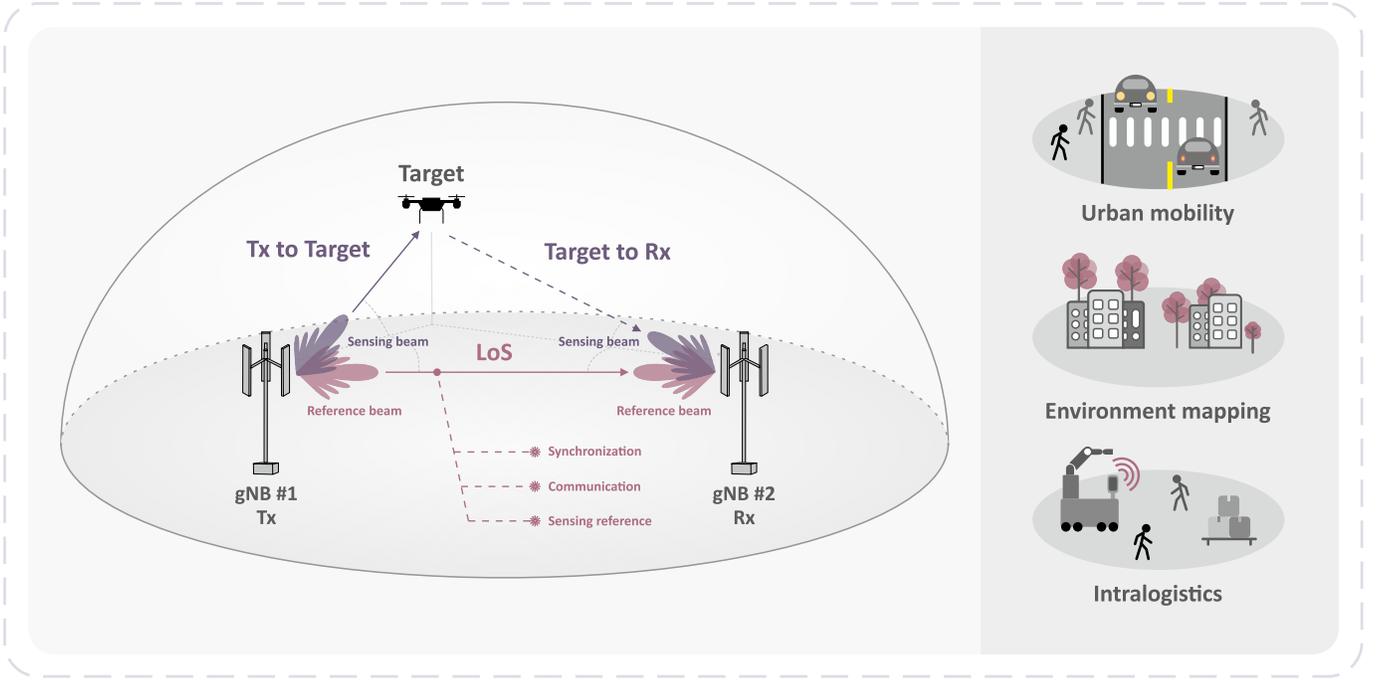}
	\captionsetup{justification=raggedright,labelsep=period,singlelinecheck=false}
	\caption{\ Conceptual representation of the proposed MIMO-OFDM-based bistatic ISAC system. W.l.o.g., it is assumed that two gNBs - one serving as a transmitter and the other as a receiver - perform beamforming to establish a LoS reference path for synchronization, communication, and radar sensing. Additionally, a secondary pair of beams is formed toward a potential target to enable sensing. Possible use cases of this system include drone detection, safety assurance in urban mobility, environmental mapping, and automation in intralogistics scenarios involving humans and autonomous guided vehicles.}\label{fig:sysModel}
	
\end{figure*}

In previous studies on bistatic \ac{SISO}-\ac{OFDM}-based \ac{ISAC} \cite{giroto2023_EuMW,brunner2024,giroto2024_tmtt}, the authors have investigated the effects and countermeasures to \ac{STO}, \ac{CFO}, and \ac{SFO}, as their estimation in \ac{ISAC} systems require much higher accuracy than in regular communication systems to ensure unbiased radar sensing. When scaling up to a \ac{MIMO} architecture, additional challenges are imposed on the \ac{ISAC} system. Focusing on practical deployments, these include, e.g., considerations of the hardware architecture and its impact on synchronization as well as the following communication and radar sensing processing chains. More specifically, challenges arise depending on how \acp{LO} and sampling clocks are distributed across the multiple transmit and receive channels, as well as on how the synchronization offset estimations are combined across each receive channel to ensure accurate sufficient coherence among the channels and therefore avoid communication and sensing performance degradation. Furthermore, aspects such as calibration, particularly when relying on \ac{OTA} synchronization, remain open challenges that must be addressed to enable effective sensing in \ac{MIMO} architectures where \ac{DoA} estimation requires phase coherence among the receive channels.

In this context, this article introduces a bistatic \ac{MIMO}-\ac{OFDM}-based \ac{ISAC} system concept, which is depicted in Fig.~\ref{fig:sysModel}. It features a multichannel transmitter with beamforming capability that generates both a reference path for synchronization, communication and sensing reference, which was assumed to be a \ac{LoS} path in previous studies \cite{giroto2023_EuMW,brunner2024,giroto2024_tmtt}, besides additional beams to enable detecting radar targets. In addition, a multichannel receiver with digital beamforming is considered. For that receiver, a distributed synchronization concept is proposed. Finally, further processing steps for communication, i.e., channel estimation, equalization, and diversity combining, and for sensing, i.e., calibration to ensure coherence among the receive channels and generation of radar images with range, Doppler shift, and \ac{DoA} information, are described. The contributions of this article can be summarized as follows:
\begin{itemize}
	\item A bistatic \ac{MIMO}-\ac{OFDM} \ac{ISAC} system concept is proposed. Besides a thorough mathematical formulation of the system model, the key processing steps at both the transmitter and the receiver side are described in detail and relevant considerations for practical deployments are made.
	
	\item A distributed synchronization concept is introduced. Assuming locally distributed \ac{LO} and sampling clocks at the transmitter and receiver, the same \ac{CFO} and \ac{SFO} are ultimately experienced at all receive channels. This allows combining local estimates at the different channels into global estimates via averaging. As for \ac{STO}, it is shown that hardware non-idealities lead to mismatches among the receive channels. To avoid \ac{ISI}, the earliest frame start point among all receive channels is taken as a global estimate and the residual \ac{STO} is later compensated at each receive channel.
	
	\item A numerical performance analysis of the proposed bistatic \ac{MIMO}-\ac{OFDM}-based \ac{ISAC} system is performed. Since only the \ac{STO} differs among the receive channels, the focus of the aforementioned analysis is placed on the robustness of both communication and radar sensing performances under uncompensated \ac{STO} mismatches, therefore allowing to predict required synchronization accuracy.
	
	\item A measurement-based validation of the \ac{MIMO}-\ac{OFDM}-based \ac{ISAC} system concept is performed with a $4\times8$ setup at $\SI{27.5}{\giga\hertz}$. The results confirm the assumption of consistent \ac{CFO} and \ac{SFO} across all receive channels, with small deviations being observed due to limited estimation accuracy. It is also confirmed that hardware non-idealities cause \ac{STO} mismatch among the channels. This can, however, be compensated with the proposed synchronization concept, ultimately ensuring sufficient coherence for \ac{DoA} estimation at the radar signal processing.
\end{itemize}

The remainder of this article is organized as follows. Section~\ref{sec:sys_model} and Section~\ref{sec:sigProc} present the system model and the processing chain, respectively, for the proposed bistatic \ac{MIMO}-\ac{OFDM}-based \ac{ISAC} system concept. Next, Section~\ref{sec:simResults} and Section~\ref{sec:measResults} present simulation and measurement results, respecitvely, besides discussing the achievable communication and radar sensing performances in the considered bistatic \ac{MIMO}-\ac{OFDM}-based \ac{ISAC} system. Finally, concluding remarks are presented in Section~\ref{sec:conclusion}.

\begin{figure}[!t]
	\centering
	
	\psfrag{d}[c][c]{$n_\mathrm{A}$}
	
	\psfrag{A}[c][c]{\small $\phi$~\textcolor[rgb]{0.7569, 0.4235, 0.5255}{$(+)$}}
	\psfrag{B}[c][c]{\small $\phi$~\textcolor[rgb]{0.7569, 0.4235, 0.5255}{$(-)$}}
	
	\psfrag{C}[c][c]{\scriptsize $\lambda_0/2$}
	
	\psfrag{N}[c][c]{\scriptsize $N_\mathrm{A}-1\phantom{/2}$}
	\psfrag{O}[c][c]{\scriptsize $N_\mathrm{A}-2\phantom{/2}$}
	\psfrag{P}[c][c]{\scriptsize $N_\mathrm{A}/2\phantom{-1}$}
	\psfrag{Q}[c][c]{\scriptsize $N_\mathrm{A}/2-1$}
	\psfrag{R}[c][c]{\scriptsize $1\phantom{N_\mathrm{A}/2-}$}
	\psfrag{S}[c][c]{\scriptsize $0\phantom{N_\mathrm{A}/2-}$}
	
	\psfrag{x}[c][c]{$x$}
	\psfrag{y}[c][c]{$y$}
	\psfrag{z}[c][c]{$z$}

	\includegraphics[width=8cm]{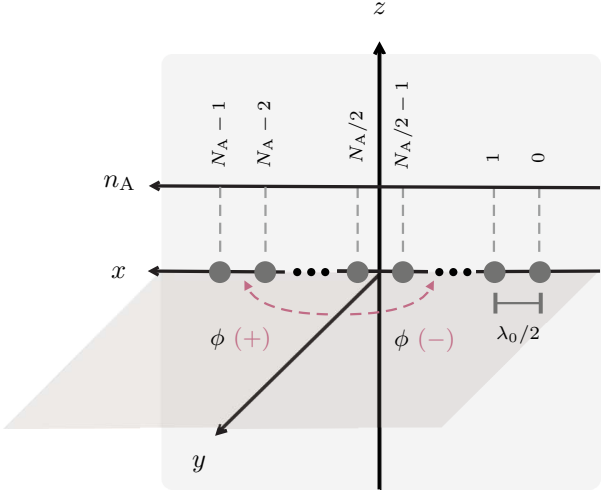}
	
	\captionsetup{justification=raggedright,labelsep=period,singlelinecheck=false}
	\caption{\ Adopted ULA arrangement and coordinate system.}\label{fig:coordRef_ULA}
\end{figure}

\section{System Model}\label{sec:sys_model}

In the considered bistatic \ac{MIMO}-\ac{OFDM}-based \ac{ISAC} system, it is assumed that both transmitting and receiving stations are static and that both their positions are known to each other. This is, e.g., the case for \acp{gNB} as depicted in Fig.~\ref{fig:sysModel}. In addition, it is w.l.o.g. assumed that a \ac{LoS} reference path is present between the stations and that both are equipped with \acp{ULA} stretching along the $x$-axis. Although these \acp{ULA} can only be used to steer beams or estimate \ac{DoA} in azimuth, it is assumed that the \ac{LoS} and any other relevant paths or radar targets are covered by the elevation beamwidth of both transmit and receive \acp{ULA}. Detailed descriptions of the antenna array geometry and transmit signal generation and propagation are provided in Sections~\ref{subsec:ula} and \ref{subsec:tx}, respectively.

\begin{figure}[!t]
	\centering
	
	\psfrag{h}[c][c]{$n_\mathrm{A}$}
	
	\psfrag{A}[c][c]{\small $\phi$~\textcolor[rgb]{0.7569, 0.4235, 0.5255}{$(-)$}}
	\psfrag{B}[c][c]{\small $\pi/2-\phi$}
	\psfrag{g}[c][c]{}
	
	\psfrag{d}[c][c]{\scriptsize $L^{2}$}
	\psfrag{e}[c][c]{\scriptsize $L^{1}$}
	\psfrag{f}[c][c]{}
	\psfrag{j}[c][c]{\scriptsize $L^\mathrm{origin}$}
	\psfrag{i}[c][c]{\scalebox{0.9}{\tiny $\Delta L^{n_\mathrm{A}}=L^{n_\mathrm{A}}-L^{\mathrm{origin}}$}}
	\psfrag{f}[c][c]{}
	
	\psfrag{C}[c][c]{\scriptsize $\lambda_0/2$}
	
	\psfrag{N}[c][c]{\scriptsize $0$}
	\psfrag{P}[c][c]{\scriptsize $1$}
	\psfrag{R}[c][c]{\scriptsize $2$}
	
	\psfrag{x}[c][c]{$x$}
	\psfrag{y}[c][c]{$y$}

	\includegraphics[width=9cm]{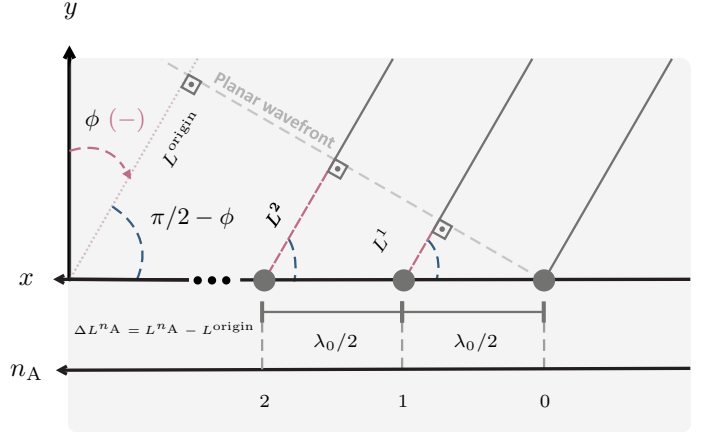}
	
	\captionsetup{justification=raggedright,labelsep=period,singlelinecheck=false}
	\caption{\ Adopted ULA arrangement and coordinate system. In this figure, the ranges $R^{n_\mathrm{A}}$ are calculated assuming that the planar wavefront crosses the $x$ axis at the position of the antenna \mbox{$n_\mathrm{A}=0$}, hence \mbox{$L^{0}=\SI{0}{\meter}$}. The length difference is then calculated as \mbox{$\Delta L^{n_\mathrm{A}}=L^{n_\mathrm{A}}-L^\mathrm{origin}$}, where $L^\mathrm{origin}$ is the range from the origin to the planar wavefront.}\label{fig:coordRef_angle}
\end{figure}

\begin{figure*}[!b]
	\hrulefill
	\vspace*{4pt}
	\setcounter{equation}{3}
	\begin{equation}\label{eq:yt}
		\resizebox{\textwidth}{!}{$
			\begin{split}
				y^{n^\mathrm{Rx}_\mathrm{A}}(t) =~ \left[ \sum_{n^\mathrm{Tx}_\mathrm{A}=0}^{N^\mathrm{Tx}_\mathrm{A}-1} \Bigg( \right. 
				& \alpha^{n^\mathrm{Tx}_\mathrm{A},n^\mathrm{Rx}_\mathrm{A}}_0~b^{\mathrm{Tx},n^\mathrm{Tx}_\mathrm{A}}_{0}~x(t-\tau^{n^\mathrm{Tx}_\mathrm{A},n^\mathrm{Rx}_\mathrm{A}}_0-\tau_\Delta)~\e^{-\im 2\pi f_\mathrm{c}\tau^{n^\mathrm{Tx}_\mathrm{A},n^\mathrm{Rx}_\mathrm{A}}_0}~\e^{\im 2\pi f_\Delta t + \psi_\Delta} \\
				& + \sum_{p=1}^{P-1} \alpha^{n^\mathrm{Tx}_\mathrm{A},n^\mathrm{Rx}_\mathrm{A}}_p~b^{\mathrm{Tx},n^\mathrm{Tx}_\mathrm{A}}_{p}~x(t-\tau^{n^\mathrm{Tx}_\mathrm{A},n^\mathrm{Rx}_\mathrm{A}}_p-\tau_\Delta)~\e^{-\im 2\pi f_\mathrm{c}\tau^{n^\mathrm{Tx}_\mathrm{A},n^\mathrm{Rx}_\mathrm{A}}_p}~\e^{\im 2\pi f^{n^\mathrm{Tx}_\mathrm{A},n^\mathrm{Rx}_\mathrm{A}}_{\mathrm{D},p}t}~\e^{\im 2\pi f_\Delta t + \psi_\Delta} \Bigg) \Bigg] \circledast h^{\mathrm{Rx},n^\mathrm{Rx}_\mathrm{A}}_\mathrm{ABE}(t)
			\end{split}$}
	\end{equation}
\end{figure*}
\setcounter{equation}{1}

\subsection{Antenna array geometry and coordinate system}\label{subsec:ula}

The  arrangement and coordinate system of the adopted \ac{ULA} for the transmitting and receiving \ac{ISAC} stations in this article are as depicted in Fig.~\ref{fig:coordRef_ULA}. This figure is composed by \mbox{$N_\mathrm{A}\in\mathbb{N}_{>0}$} antenna elements labeled as \mbox{$n_\mathrm{A}\in\{0,1,\dots,N_\mathrm{A}-1\}$}. All antenna elements are placed along the $x$ axis with spacing \mbox{$\lambda_0/2$} among consecutive elements, where \mbox{$\lambda_0=c_0/f_\text{c}$} denotes the corresponding wavelength to the carrier frequency $f_\text{c}$ and $c_0$ is the speed of light in vacuum. In addition, it is assumed that they face the positive $y$ direction. The $x$, $y$, and $z$ coordinates of the $n_\mathrm{A}\mathrm{th}$ antenna element are denoted by the position vector \mbox{$\bm{\rho}^{n_\mathrm{A}}\in\mathbb{R}^{3\times 1}|\bm{\rho}^{n_\mathrm{A}}=\left[\rho^{n_\mathrm{A}}_x,\rho^{n_\mathrm{A}}_y,\rho^{n_\mathrm{A}}_z\right]^T$} such that \mbox{$\rho^{n_\mathrm{A}}_x=(\lambda_0/2)[(-N_\mathrm{A}+1)/2+n_\mathrm{A}]$} and \mbox{$\rho^{n_\mathrm{A}}_y=\rho^{n_\mathrm{A}}_z=0~\forall~n_\mathrm{A}$}. An azimuth angle $\phi\in[-\pi/2,\pi/2)$, which can either be a \ac{DoD} in the case of a transmit \ac{ULA} or \ac{DoA} in the case of a receive \ac{ULA}, is also defined in Fig.~\ref{fig:coordRef_ULA}. Positive $\phi$ angles, i.e., $\phi\in[0,\pi/2)$, are measured if they correspond to directions between positive $x$ and $y$ axes, while negative $\phi$ angles, i.e., $\phi\in[-\pi/2,0)$, are measured for directions between negative $x$ and positive $y$ axes. If signals are either beamformed torwards or received from an angle $\phi$, a different electromagnetic wave propagation length will be traveled by the signals associated with each of the $N_\mathrm{A}$ antennas. Based on Fig.~\ref{fig:coordRef_angle}, the length difference $\Delta L^{n_\mathrm{A}}$ for the $n_\mathrm{A}\mathrm{th}$ antenna w.r.t. the origin of the array at \mbox{$(x,y,z)=(0,0,0)$} is given by
\setcounter{equation}{0}
\begin{align}\label{eq:rDelta}
	\Delta L^{n_\mathrm{A}} &= -\left(\lambda_0/2\right)[(-N_\mathrm{A}+1)/2+n_\mathrm{A}]\cos\left(\pi/2-\phi\right)\nonumber\\
	&= -\left(\lambda_0/2\right)[(-N_\mathrm{A}+1)/2+n_\mathrm{A}]\sin\left(\phi\right).
\end{align}

\subsection{Transmit signal generation and propagation}\label{subsec:tx}

At the transmitter side, a transmit signal \mbox{$x(t)\in\mathbb{C}$} that occupies a bandwidth \mbox{$B\leq F_\mathrm{s}$}, where $F_\mathrm{s}$ is the sampling frequency associated with sampling period \mbox{$T_\mathrm{s}=1/F_\mathrm{s}$}, is generated. Next, $x(t)$ is up-converted to the carrier frequency \mbox{$f_\text{c}\gg B$} with an \ac{I/Q} mixer, and beamforming is performed on the resulting signal towards \mbox{$P\in\mathbb{N}_{>0}$} steering directions.  Neglecting the beam sidelobes for simplicity, this ultimately results in $P$ propagation paths labeled as \mbox{$p\in\{0,1,\dots,P-1\}$} towards each out of the \mbox{$N^\mathrm{Rx}_\mathrm{A}\in\mathbb{N}_{\geq0}$} receive antennas also arranged as depicted in Fig.~\ref{fig:coordRef_ULA}. The first, \mbox{$p=0$}, is the \ac{LoS} direction towards the receiver. The other \mbox{$P-1$} beams, i.e., \mbox{$p=1$} to \mbox{$p=P-1$}, are steered in strategic directions to perform sensing and possibly serve other communication receivers. Every $p\mathrm{th}$ path is associated with an attenuation factor $\alpha^{n^\mathrm{Tx}_\mathrm{A},n^\mathrm{Rx}_\mathrm{A}}_p$, a delay $\tau^{n^\mathrm{Tx}_\mathrm{A},n^\mathrm{Rx}_\mathrm{A}}_p$, an a Doppler shift $f^{n^\mathrm{Tx}_\mathrm{A},n^\mathrm{Rx}_\mathrm{A}}_{\mathrm{D},p}$, with \mbox{$n^\mathrm{Rx}_\mathrm{A}\in\{0,1,\dots,N^\mathrm{Rx}_\mathrm{A}-1\}$} denoting the receive antenna index. Furthermore, it is assumed that the $p\mathrm{th}$ path is associated with an ideal azimuth \ac{DoD} $\phi^\mathrm{Tx}_{p}$, which corresponds to the true transmitter-referred direction of the receiver for \mbox{$p=0$} and the true direction of radar targets for \mbox{$p\in\{1,\cdots,P-1\}$}.

The described transmit beamforming is achieved by transmitting weighted copies of the same up-converted version of $x(t)$ through each of the \mbox{$N^\mathrm{Tx}_\mathrm{A}\in\mathbb{N}_{>0}$} transmit channels, which are assumed to be associated to their own antenna. The weight at each $n^\mathrm{Tx}_\mathrm{A}\mathrm{th}$ channel corresponds to the sum of the $n^\mathrm{Tx}_\mathrm{A}\mathrm{th}$ element of the transmit beamsteering vectors \mbox{$\mathbf{b}^{\mathrm{Tx}}_{p}\in\mathbb{C}^{N^\mathrm{Tx}_\mathrm{A}\times 1}$} associated with steering directions of index $p$ over all $P$ steering directions. The $n^\mathrm{Tx}_\mathrm{A}\mathrm{th}$ element of $\mathbf{b}^{\mathrm{Tx}}_{p}$ is given by
\begin{equation}\label{eq:txBeamfW}
	b^{\mathrm{Tx},n^\mathrm{Tx}_\mathrm{A}}_{p} = \e^{-\im 2\pi\left(\rho^{\mathrm{Tx},n^\mathrm{Tx}_\mathrm{A}}_x/\lambda_0\right)\sin\left(\widehat{\phi^\mathrm{Tx}_{p}}\right)}.
\end{equation}
In this equation, $\rho^{\mathrm{Tx},n^\mathrm{Tx}_\mathrm{A}}_x$ is the $x$ coordinate of the \mbox{$n^\mathrm{Tx}_\mathrm{A}\mathrm{th}$} transmit antenna, whose position is represented by the vector \mbox{$\bm{\rho}^{\mathrm{Tx},n^\mathrm{Tx}_\mathrm{A}}\in\mathbb{R}^{3\times 1}|\bm{\rho}^{\mathrm{Tx},n^\mathrm{Tx}_\mathrm{A}}=\left[\rho^{\mathrm{Tx},n^\mathrm{Tx}_\mathrm{A}}_x,\rho^{\mathrm{Tx},n^\mathrm{Tx}_\mathrm{A}}_y,\rho^{\mathrm{Tx},n^\mathrm{Tx}_\mathrm{A}}_z\right]^T$}. As the model from Fig.~\ref{fig:coordRef_ULA} is considered, it holds that
\begin{equation}\label{eq:pos_n_Tx}
	\rho^{\mathrm{Tx},n^\mathrm{Tx}_\mathrm{A}}_x=(\lambda_0/2)[(-N^\mathrm{Tx}_\mathrm{A}+1)/2+n^\mathrm{Tx}_\mathrm{A}]
\end{equation}
as well as \mbox{$\rho^{\mathrm{Tx},n^\mathrm{Tx}_\mathrm{A}}_y=\rho^{\mathrm{Tx},n^\mathrm{Tx}_\mathrm{A}}_z=0~\forall~n^\mathrm{Tx}_\mathrm{A}$}. In addition, $\widehat{\phi^\mathrm{Tx}_{p}}$ is the estimate of $\phi^\mathrm{Tx}_{p}$ that is used as the azimuth \ac{DoD} of the $p\mathrm{th}$ steering direction.

\begin{figure*}[!b]
	\hrulefill
	\vspace*{4pt}
	\setcounter{equation}{14}
	\begin{equation}\label{eq:yt2}
		\resizebox{\textwidth}{!}{$
			\begin{split}
				y^{n^\mathrm{Rx}_\mathrm{A}}(t) =~ \Bigg(&\alpha^{n^\mathrm{Tx}_\mathrm{A},n^\mathrm{Rx}_\mathrm{A}}_0~N^\mathrm{Tx}_\mathrm{A}~x(t-\tau_0-\tau_\Delta)~\e^{-\im 2\pi f_\mathrm{c}\tau^{\mathrm{Tx}}_0}~\e^{-\im 2\pi f_\mathrm{c}\tau^{\mathrm{Rx},n^\mathrm{Rx}_\mathrm{A}}_0}~\e^{\im 2\pi f_\Delta t + \psi_\Delta}\\
				&+\sum_{p=1}^{P-1}\alpha^{n^\mathrm{Tx}_\mathrm{A},n^\mathrm{Rx}_\mathrm{A}}_p~N^\mathrm{Tx}_\mathrm{A}~x(t-\tau_p-\tau_\Delta)~\e^{-\im 2\pi f_\mathrm{c}\tau^{\mathrm{Tx}}_p}~\e^{-\im 2\pi f_\mathrm{c}\tau^{\mathrm{Rx},n^\mathrm{Rx}_\mathrm{A}}_p}~\e^{\im 2\pi f^{n^\mathrm{Tx}_\mathrm{A},n^\mathrm{Rx}_\mathrm{A}}_{\mathrm{D},p}t}~\e^{\im 2\pi f_\Delta t + \psi_\Delta}\Bigg) \circledast h^{\mathrm{Rx},n^\mathrm{Rx}_\mathrm{A}}_\mathrm{ABE}(t)
			\end{split}$}
	\end{equation}
\end{figure*}

After propagation through the aforementioned $P$ paths, attenuated, delayed and Doppler-shifted versions of the up-converted and beamformed version of $x(t)$ are captured by the \mbox{$n^\mathrm{Rx}_\mathrm{A}\mathrm{th}$} receive antenna. This constitutes the \ac{BB} signal at the \mbox{$n^\mathrm{Rx}_\mathrm{A}\mathrm{th}$} receive channel, which after down-conversion into the baseband with an \ac{I/Q} mixer is denoted as \mbox{$y^{n^\mathrm{Rx}_\mathrm{A}}(t)\in\mathbb{C}$} and expressed as in \eqref{eq:yt}. In this equation, $\alpha^{n^\mathrm{Tx}_\mathrm{A},n^\mathrm{Rx}_\mathrm{A}}_p$, $\tau^{n^\mathrm{Tx}_\mathrm{A},n^\mathrm{Rx}_\mathrm{A}}_p$, and $f^{n^\mathrm{Tx}_\mathrm{A},n^\mathrm{Rx}_\mathrm{A}}_{\mathrm{D},p}$ are the attenuation factor, delay, and Doppler shift associated with the $p\mathrm{th}$ path between the \mbox{$n^\mathrm{Tx}_\mathrm{A}\mathrm{th}$} transmit and the \mbox{$n^\mathrm{Rx}_\mathrm{A}\mathrm{th}$} receive antennas. Note that \mbox{$f^{n^\mathrm{Tx}_\mathrm{A},n^\mathrm{Rx}_\mathrm{A}}_{\mathrm{D},0}=\SI{0}{\hertz}$} since transmitting and receiving stations are assumed to be static and there is therefore no Doppler shift for the \ac{LoS} path between them labeled as $p=0$. The propagation delays $\tau^{n^\mathrm{Tx}_\mathrm{A},n^\mathrm{Rx}_\mathrm{A}}_p$ also result in phase shifts, which are denoted by $\e^{-\im 2\pi f_\mathrm{c}\tau^{n^\mathrm{Tx}_\mathrm{A},n^\mathrm{Rx}_\mathrm{A}}_p}$. Time and frequency offsets between transmitter and receiver are also considered in \eqref{eq:yt}. These are namely the \ac{STO} $\tau_\Delta$, as well as the \ac{CFO} $f_\Delta$ and its resulting phase rotation $\psi_\Delta$. Finally, \mbox{$h^{\mathrm{Rx},n^\mathrm{Rx}_\mathrm{A}}_\mathrm{ABE}(t)\in\mathbb{C}$} is the \ac{ABE} \ac{CIR} of the \mbox{$n^\mathrm{Rx}_\mathrm{A}\mathrm{th}$} receive channel associated with a corresponding \ac{CFR} \mbox{$H^{n^\mathrm{Rx}_\mathrm{A}}_{\mathrm{ABE}}(f)\in\mathbb{C}$}. They account for hardware non-idealities that may lead to slightly different delays and therefore phases among the $N^\mathrm{Rx}_\mathrm{A}$ receive channels. It is henceforth assumed that the aforementioned \ac{CIR} assumed have a dominant path with delay $\tau_\mathrm{ABE}^{n^\mathrm{Rx}_\mathrm{A}}$.

\section{Transmit and Receive Signal Processing}\label{sec:sigProc}

Based on the system model outlined in Section~\ref{sec:sys_model}, the transmit and receive signal processing steps in the proposed bistatic \ac{MIMO}-\ac{OFDM}-based \ac{ISAC} system are discussed in this section. More specifically, Section~\ref{subsec:bfTx} discusses the choice of the transmit beamforming weights, while Section~\ref{subsec:rxProc} discusses the receive signal processing chain for both communication and bistatic radar sensing. 

\subsection{Choice of transmit beamforming weights}\label{subsec:bfTx}

Based on \eqref{eq:rDelta} and the discussion on it in Section~\ref{subsec:ula}, it is known that a electromagnetic wave propagation length difference $\Delta L_{p}^{n_\mathrm{Tx}}$ for the $n_\mathrm{Tx}\mathrm{th}$ antenna w.r.t. the origin of the transmit array at \mbox{$(x^\mathrm{Tx},y^\mathrm{Tx},z^\mathrm{Tx})=(0,0,0)$} given by
\setcounter{equation}{4}
\begin{equation}\label{eq:rDeltaTx}
	\Delta L_{p}^{n_\mathrm{Tx}} = -(\lambda_0/2)[(-N^\mathrm{Tx}_\mathrm{A}+1)/2+n^\mathrm{Tx}_\mathrm{A}]\sin\left(\phi^\mathrm{Tx}_{p}\right)	
\end{equation}
will be experienced. Converting this length difference into delay yields \mbox{$\tau_{\Delta,p}^{n_\mathrm{Tx}}=\Delta L_{p}^{n_\mathrm{Tx}}/c_0$}, which can be alternatively expressed as
\begin{equation}\label{eq:tauDeltaTx}
	\tau_{\Delta,p}^{n_\mathrm{Tx}} = -\left(2f_\mathrm{c}\right)^{-1}[(-N^\mathrm{Tx}_\mathrm{A}+1)/2+n^\mathrm{Tx}_\mathrm{A}]\sin\left(\phi^\mathrm{Tx}_{p}\right).
\end{equation}
Next, the propagation delays $\tau^{n^\mathrm{Tx}_\mathrm{A},n^\mathrm{Rx}_\mathrm{A}}_p$ are denoted as 
\begin{equation}
	\tau^{n^\mathrm{Tx}_\mathrm{A},n^\mathrm{Rx}_\mathrm{A}}_p = \tau^{\mathrm{Tx},n^\mathrm{Tx}_\mathrm{A}}_p+\tau^{\mathrm{Rx},n^\mathrm{Rx}_\mathrm{A}}_p,
\end{equation}
where, for $p\in\{1,\dots,P-1\}$, $\tau^{\mathrm{Tx},n^\mathrm{Tx}_\mathrm{A}}_p$ is the propagation delay between the \mbox{$n^\mathrm{Tx}_\mathrm{A}\mathrm{th}$} transmit antenna and the radar target and $\tau^{\mathrm{Rx},n^\mathrm{Rx}_\mathrm{A}}_p$ is the propagation delay between the radar target and the \mbox{$n^\mathrm{Rx}_\mathrm{A}\mathrm{th}$} receive antenna, both for the \mbox{$p\mathrm{th}$} path. For the \ac{LoS} path, i.e., \mbox{$p=0$}, it is assumed that \mbox{$\tau^{\mathrm{Tx},n^\mathrm{Tx}_\mathrm{A}}_0=\tau^{n^\mathrm{Tx}_\mathrm{A},n^\mathrm{Rx}_\mathrm{A}}_0$} and \mbox{$\tau^{\mathrm{Rx},n^\mathrm{Rx}_\mathrm{A}}_0=\SI{0}{\second}$}. Using the result from \eqref{eq:tauDeltaTx}, the propagation delays $\tau^{\mathrm{Tx},n^\mathrm{Tx}_\mathrm{A}}_p$ are further expanded as
\begin{equation}\label{eq:tauTxPlusDelta}
	\tau^{\mathrm{Tx},n^\mathrm{Tx}_\mathrm{A}}_p = \tau^{\mathrm{Tx}}_p +\tau_{\Delta,p}^{n_\mathrm{Tx}},
\end{equation}
where
\begin{equation}
	\tau^{\mathrm{Tx}}_p = R^{\mathrm{Tx}}_p/c_0
\end{equation}
and $R^{\mathrm{Tx}}_p$ denotes the range of the $p\mathrm{th}$ target w.r.t. the origin of the transmit array at \mbox{$(x^\mathrm{Tx},y^\mathrm{Tx},z^\mathrm{Tx})=(0,0,0)$}. The results from equations \eqref{eq:rDeltaTx} to \eqref{eq:tauTxPlusDelta} allow expanding the expression of the delay-induced phase shifts $\e^{-\im 2\pi f_\mathrm{c}\tau^{n^\mathrm{Tx}_\mathrm{A},n^\mathrm{Rx}_\mathrm{A}}_p}$ as
\begin{equation}\label{eq:phaseShiftTx0}
	\left(\e^{-\im 2\pi f_\mathrm{c}\tau^{\mathrm{Tx}}_p}\e^{-\im 2\pi f_\mathrm{c}\tau^{\mathrm{Rx},n^\mathrm{Rx}_\mathrm{A}}_p}\right)\e^{-\im 2\pi f_\mathrm{c}\tau_{\Delta,p}^{n_\mathrm{Tx}}},
\end{equation}
or further as
\begin{equation}\label{eq:phaseShiftTx}
	\left(\e^{-\im 2\pi f_\mathrm{c}\tau^{\mathrm{Tx}}_p}\e^{-\im 2\pi f_\mathrm{c}\tau^{\mathrm{Rx},n^\mathrm{Rx}_\mathrm{A}}_p}\right)\e^{\im \pi [(-N^\mathrm{Tx}_\mathrm{A}+1)/2+n^\mathrm{Tx}_\mathrm{A}]\sin\left(\phi^\mathrm{Tx}_{p}\right)}.
\end{equation}

To ensure maximum transmit beamforming gain, the weights $b^{\mathrm{Tx},n^\mathrm{Tx}_\mathrm{A}}_{p}$ must be chosen such that $\widehat{\phi^\mathrm{Tx}_{p}}=\phi^\mathrm{Tx}_{p}$. This can be achieved by setting the arguments of the exponential function defining the beamforming weights in \eqref{eq:txBeamfW}, equal to the negative of the arguments of the rightmost exponential function in \eqref{eq:phaseShiftTx}, i.e., 
\begin{equation}\label{eq:equalExp}
	\resizebox{\columnwidth}{!}{$\im 2\pi\left(\rho^{\mathrm{Tx},n^\mathrm{Tx}_\mathrm{A}}_x/\lambda_0\right)\sin\left(\widehat{\phi^\mathrm{Tx}_{p}}\right) = \im \pi [(-N^\mathrm{Tx}_\mathrm{A}+1)/2+n^\mathrm{Tx}_\mathrm{A}]\sin\left(\phi^\mathrm{Tx}_{p}\right).$}
\end{equation}
Rearranging the expression in \eqref{eq:equalExp} and knowing from \eqref{eq:pos_n_Tx} that \mbox{$n^\mathrm{Tx}_\mathrm{A}=\rho^{\mathrm{Tx},n^\mathrm{Tx}_\mathrm{A}}_x/\left(\lambda_0/2\right)-(-N^\mathrm{Tx}_\mathrm{A}+1)/2$}, \mbox{$\widehat{\phi^\mathrm{Tx}_{p}}=\phi^\mathrm{Tx}_{p}$} is obtained, which implies that the maximum transmit beamforming gains are obtained if the steering angles $\widehat{\phi^\mathrm{Tx}_{p}}$ are set equal to the angles $\phi^\mathrm{Tx}_{p}$ corresponding to the \ac{LoS} path ($p=0$) or the radar targets ($p \in {1, \dots, P-1}$). This ultimately results in
\begin{equation}\label{eq:sum_B}
	\resizebox{\hsize}{!}{$\sum_{n^\mathrm{Tx}_\mathrm{A}=0}^{N^\mathrm{Tx}_\mathrm{A}-1} b^{\mathrm{Tx},n^\mathrm{Tx}_\mathrm{A}}_{p} \e^{-\im 2\pi f_\mathrm{c} \tau^{n^\mathrm{Tx}_\mathrm{A},n^\mathrm{Rx}_\mathrm{A}}_p} = N^\mathrm{Tx}_\mathrm{A}\left(\e^{-\im 2\pi f_\mathrm{c}\tau^{\mathrm{Tx}}_p}\e^{-\im 2\pi f_\mathrm{c}\tau^{\mathrm{Rx},n^\mathrm{Rx}_\mathrm{A}}_p}\right)$.}
\end{equation}
It is then considered that the delays for every \mbox{$p\mathrm{th}$} path are approximately the same for every combination of transmit and receive channels, i.e.,
\begin{equation}\label{eq:delay_p}
	\tau^{n^\mathrm{Tx}_\mathrm{A},n^\mathrm{Rx}_\mathrm{A}}_p\approx\tau_p~\quad\forall~n^\mathrm{Tx}_\mathrm{A},n^\mathrm{Rx}_\mathrm{A},p.
\end{equation}
This assumption is reasonable since the phase rotations induced by the slight difference in the delays are still considered. Based on \eqref{eq:sum_B} and \eqref{eq:delay_p}, \eqref{eq:yt} can be rewritten as \eqref{eq:yt2}.

It is worth highlighting that, in practical deployments, the $N^\mathrm{Tx}_\mathrm{A}$ transmit channels of the \acp{AFE} will likely have different \acp{CFR} \mbox{$H^{n^\mathrm{Tx}_\mathrm{A}}_{\mathrm{AFE}}(f)\in\mathbb{C}$}. These must be estimated beforehand with calibration measurements and compensated via \ac{DPD} to ensure that transmit beamforming is correctly performed.

\subsection{Receive signal processing chain}\label{subsec:rxProc}

After being sampled at each \mbox{$n^\mathrm{Rx}_\mathrm{A}\mathrm{th}$} receive channel, the discrete-time domain equivalent sequences to the baseband signals $y^{n^\mathrm{Rx}_\mathrm{A}}(t)$ must undergo synchronization so that they can be undergo communication processing to allow estimating the transmit \ac{OFDM} frame which will then be used for bistatic radar signal processing as described in \cite{giroto2023_EuMW,brunner2024}. The three aforementioned processing steps are described in further detail as follows.

\begin{figure*}[!b]
	\hrulefill
	\vspace*{4pt}
	\setcounter{equation}{20}
	\begin{equation}\label{eq:ys}
		\resizebox{\textwidth}{!}{$
			\begin{split}
				y^{n^\mathrm{Rx}_\mathrm{A}}_s \approx~ \left[\Bigg(\right.&\alpha^{n^\mathrm{Tx}_\mathrm{A},n^\mathrm{Rx}_\mathrm{A}}_0~N^\mathrm{Tx}_\mathrm{A}~x(t)~\e^{-\im 2\pi f_\mathrm{c}\tau^{\mathrm{Tx}}_0}~\e^{-\im 2\pi f_\mathrm{c}\tau^{\mathrm{Rx},n^\mathrm{Rx}_\mathrm{A}}_0}~\e^{\im\psi_\Delta}\\ &\left.+\sum_{p=1}^{P-1}\alpha^{n^\mathrm{Tx}_\mathrm{A},n^\mathrm{Rx}_\mathrm{A}}_p~N^\mathrm{Tx}_\mathrm{A}~x\left(t-(\tau_p-\tau_0)\right)~\e^{-\im 2\pi f_\mathrm{c}\tau^{\mathrm{Tx}}_p}~\e^{-\im 2\pi f_\mathrm{c}\tau^{\mathrm{Rx},n^\mathrm{Rx}_\mathrm{A}}_p}~\e^{\im 2\pi f^{n^\mathrm{Tx}_\mathrm{A},n^\mathrm{Rx}_\mathrm{A}}_{\mathrm{D},p}t}~\e^{\im\psi_\Delta}\left.\Bigg)\circledast h^{\mathrm{Rx},n^\mathrm{Rx}_\mathrm{A}}_\mathrm{ABE}\left(t-\left[\widehat{\tau}_\Delta^\mathrm{MIMO}-\left(\tau_0+\tau_\Delta\right)\right]\right)\right]~\right\rvert_{t=sT_\mathrm{s}}\\\phantom{.}
			\end{split}$}
	\end{equation}
\end{figure*}

\subsubsection{Synchronization}\label{subsubsec:sync}

In this article, is assumed that \ac{STO} and \ac{CFO} are individually estimated for each receive channel based on what is described for the \ac{SISO} case in \cite{giroto2023_EuMW}. More specifically, a first, coarse estimation of the \ac{CFO} is obtained at each receive channel with the \ac{SC} algorithm. After correcting the \ac{CFO}, an \ac{STO} estimate for each receive channel is obtained via cross-correlation with a preamble \ac{OFDM} symbol to yield a more accurate frame start point estimate than what can be achieved with the \ac{SC} algorithm. Next, the \ac{SFO} is also estimated for each receive channel with the \ac{TITO} algorithm proposed in \cite{giroto2024_tmtt}. After resampling based on the aforementioned \ac{SFO} estimate, a fine-tuning of the \ac{STO} and \ac{CFO} estimates is performed based on pilot subcarriers.

Since a \ac{MIMO} architecture is assumed, the \ac{CFO} and \ac{SFO} estimates are averaged to yield improved accuracy before these offsets can be corrected in the proposed bistatic \ac{MIMO}-\ac{OFDM}-based \ac{ISAC} system. This can be done since there are two groups of \acp{LO} and sampling clocks, one shared by all transmit channels and another one shared by all receive channels. Consequently, the same \ac{CFO} and \ac{SFO} must be experienced by all $N^\mathrm{Rx}_\mathrm{A}$ receive channels. The \ac{CFO} estimate $\widehat{f}_\mathrm{\Delta}^\mathrm{MIMO}$ and the estimate $\widehat{\delta}^\mathrm{MIMO}$ of the normalized \ac{SFO} by the sampling frequency $T_\mathrm{s}$ \cite{giroto2024_tmtt} in the considered bistatic \ac{MIMO}-\ac{OFDM}-based \ac{ISAC} system are therefore obtained as
\setcounter{equation}{15}
\begin{equation}
	\widehat{f}_\mathrm{\Delta}^\mathrm{MIMO} = \frac{1}{N^\mathrm{Rx}_\mathrm{A}}\sum_{n^\mathrm{Rx}_\mathrm{A}=0}^{N^\mathrm{Rx}_\mathrm{A}-1}\widehat{f}_\mathrm{\Delta}^{n^\mathrm{Rx}_\mathrm{A}}
\end{equation}
and
\begin{equation}
	\widehat{\delta}^\mathrm{MIMO} = \frac{1}{N^\mathrm{Rx}_\mathrm{A}}\sum_{n^\mathrm{Rx}_\mathrm{A}=0}^{N^\mathrm{Rx}_\mathrm{A}-1}\widehat{\delta}^{n^\mathrm{Rx}_\mathrm{A}},
\end{equation}
where $\widehat{f}_\mathrm{\Delta}^{n^\mathrm{Rx}_\mathrm{A}}$ and $\widehat{\delta}^{n^\mathrm{Rx}_\mathrm{A}}$ are the \ac{CFO} and normalized \ac{SFO} estimates for the \mbox{$n^\mathrm{Rx}_\mathrm{A}\mathrm{th}$} receive channel, respectively.

As for the \ac{STO}, the same cannot be done since different delays are experienced as each \mbox{$n^\mathrm{Rx}_\mathrm{A}\mathrm{th}$} receive channel is associated with its own \ac{ABE} \ac{CIR} \mbox{$h^{\mathrm{Rx},n^\mathrm{Rx}_\mathrm{A}}_\mathrm{ABE}(t)$}. Although \mbox{$h^{\mathrm{Rx},n^\mathrm{Rx}_\mathrm{A}}_\mathrm{ABE}(t)$} tends to have a dominant tap as the connections between the \ac{ABE} elements are matched, the experienced delays may differ among the $N^\mathrm{Rx}_\mathrm{A}$ even if only by a fraction of the sampling period $T_\mathrm{s}$. This is, however, enough to break the phase coherence among receive channels and impair the \ac{DoA} estimation processing performed as later explained in Section~\ref{subsubsec:bistRadarSP}. In this sense, the earliest estimated \ac{OFDM} frame start point among all $N^\mathrm{Rx}_\mathrm{A}$ receive channels is taken as a common start point for all channels to ensure that none of them is impaired by \ac{ISI}. The \ac{STO} estimate $\widehat{\tau}_\Delta^\mathrm{MIMO}$ in the considered bistatic \ac{MIMO}-\ac{OFDM}-based \ac{ISAC} system is therefore calculated as
\begin{equation}\label{eq:STO_est1}
	\widehat{\tau}_\Delta^\mathrm{MIMO} = \min_{n^\mathrm{Rx}_\mathrm{A} \in \{0, \dots, N^\mathrm{Rx}_\mathrm{A} - 1\}} \widehat{\tau}_\Delta^{n^\mathrm{Rx}_\mathrm{A}},
\end{equation}
where $\widehat{\tau}_\Delta^{n^\mathrm{Rx}_\mathrm{A}}$ is the \ac{STO} estimate for the \mbox{$n^\mathrm{Rx}_\mathrm{A}\mathrm{th}$} receive channel. It is worth highlighting that the \ac{STO} estimate will not only include the \ac{STO} itself, but also the delay of the \ac{LoS} reference path as they are not distinguishable at the receiver side. Disregarding estimation bias for the sake of simplicity, the \ac{STO} estimate $\widehat{\tau}_\Delta^{n^\mathrm{Rx}_\mathrm{A}}$ at the \mbox{$n^\mathrm{Rx}_\mathrm{A}\mathrm{th}$} receive channel will ideally be given by
\begin{equation}\label{eq:STO_channel}
	\widehat{\tau}_\Delta^{n^\mathrm{Rx}_\mathrm{A}} = \tau_0 + \tau_\Delta + \tau_\mathrm{ABE}^{n^\mathrm{Rx}_\mathrm{A}}.
\end{equation}
In other words, $\widehat{\tau}_\Delta^{n^\mathrm{Rx}_\mathrm{A}}$ will be ideally the result of the sum of the delay associated with the \ac{LoS} reference path, the actual \ac{STO} between transmitter and receiver, and the delay of the dominant path of the \ac{ABE} \ac{CIR} at the \mbox{$n^\mathrm{Rx}_\mathrm{A}\mathrm{th}$} receive channel. Consequently, \eqref{eq:STO_est1} can be rewritten as
\begin{equation}\label{eq:STO_est2}
	\widehat{\tau}_\Delta^\mathrm{MIMO} = \tau_0 + \tau_\Delta + \min_{n^\mathrm{Rx}_\mathrm{A} \in \{0, \dots, N^\mathrm{Rx}_\mathrm{A} - 1\}} \tau_\mathrm{ABE}^{n^\mathrm{Rx}_\mathrm{A}}
\end{equation}
if the assumption of the absence of estimation bias is kept.

Once the start point of the \ac{OFDM} frame has been defined based on the \ac{STO} estimate $\widehat{\tau}_\Delta^\mathrm{MIMO}$, the normalized \ac{SFO} estimate $\widehat{\delta}^\mathrm{MIMO}$ is used to perform resampling of the receive signal, followed by a \ac{CFO} correction based on $\widehat{f}_\mathrm{\Delta}^\mathrm{MIMO}$. Since the \ac{STO} was corrected only at the sample level and the \ac{ABE} \ac{CIR} delays $\tau_\mathrm{ABE}^{n^\mathrm{Rx}_\mathrm{A}}$ are different among the $N^\mathrm{Rx}_\mathrm{A}$ receive channels, a fine tuning is required. Available pilot subcarriers in the \ac{OFDM} frame are then used to estimate the residual \ac{STO} and \ac{CFO} associated with the \ac{LoS} reference path, consequently allowing their correction and aligning the \ac{LoS} reference path at a delay of $\SI{0}{\second}$ and a Doppler shift of $\SI{0}{\hertz}$ throughout all $N^\mathrm{Rx}_\mathrm{A}$ receive channels.

The described synchronization for the proposed bistatic \ac{MIMO}-\ac{OFDM}-based \ac{ISAC} system results in a discrete-time domain sequence $y^{n^\mathrm{Rx}_\mathrm{A}}_s\in\mathbb{C}$ for every \mbox{$n^\mathrm{Rx}_\mathrm{A}\mathrm{th}$} receive channel, which is expressed as in \eqref{eq:ys}. At the \mbox{$n^\mathrm{Rx}_\mathrm{A}\mathrm{th}$} receive channel, this sequence undergoes \ac{S/P} conversion to yield a discrete-time domain \ac{OFDM} frame. After \ac{CP} removal, the \ac{OFDM} symbols in the aforementioned frame undergo \ac{DFT} resulting in the discrete-frequency domain \ac{OFDM} frame \mbox{$\mathbf{Y}^{n^\mathrm{Rx}_\mathrm{A}}\in\mathbb{C}^{N\times M}$} with \mbox{$N\in\mathbb{N}_{\geq 0}$} subcarriers and \mbox{$M\in\mathbb{N}_{\geq 0}$} \ac{OFDM} symbols. From this frame, pilots can be extracted to perform the aforementioned residual \ac{STO} and \ac{CFO} corrections before further communication and bistatic radar signal processing take place.

\subsubsection{Communication signal processing}\label{subsubsec:commSP}

The communication signal processing for the \mbox{$n^\mathrm{Rx}_\mathrm{A}\mathrm{th}$} receive channel starts with channel estimation. This is performed based on pilot subcarriers, which are assumed to be distributed in the \ac{OFDM} frame with a regular spacing of \mbox{$\Delta N_\mathrm{pil}\in\mathbb{N}_{>0}$} subcarriers and \mbox{$\Delta M_\mathrm{pil}\in\mathbb{N}_{>0}$} \ac{OFDM} symbols, resulting in a total of \mbox{$M_\mathrm{pil}=M/\Delta M_\mathrm{pil}$} pilot \ac{OFDM} symbols, each with \mbox{$N_\mathrm{pil}=N/\Delta N_\mathrm{pil}$} pilot subcarriers \cite{giroto2024_tmtt}. The estimated communication \acp{CFR} at the positions of the aforementioned pilot subcarriers then undergo two-dimensional interpolation to yield the communication \ac{CFR} matrix \mbox{$\mathbf{H}_\mathrm{CFR}^{n^\mathrm{Rx}_\mathrm{A}}\in\mathbb{C}^{N\times M}$}.

After channel estimation, the discrete-frequency domain \ac{OFDM} frames $\mathbf{Y}^{n^\mathrm{Rx}_\mathrm{A}}$ at all \mbox{$N^\mathrm{Rx}_\mathrm{A}$} receive channels undergo \ac{MRC} \cite{kahn1954,brennan1959} as an additional step to what is performed in the \ac{SISO} case in \cite{giroto2023_EuMW}. This results in a single discrete-frequency domain \ac{OFDM} frame \ac{OFDM} \mbox{$\mathbf{Y}^\mathrm{MRC}\in\mathbb{C}^{N\times M}$}, whose element at its \mbox{$n\mathrm{th}$} row, \mbox{$n\in\{0,1,\dots,N-1\}$}, and \mbox{$m\mathrm{th}$} column, \mbox{$m\in\{0,1,\dots,M-1\}$}, is given by
\setcounter{equation}{21}
\begin{equation}\label{eq:MRC}
	Y^\mathrm{MRC}_{n,m} = \frac{\sum_{n^\mathrm{Rx}_\mathrm{A}=0}^{N^\mathrm{Rx}_\mathrm{A}-1}\left[Y^{n^\mathrm{Rx}_\mathrm{A}}_{n,m}~\left(H^{n^\mathrm{Rx}_\mathrm{A}}_{\mathrm{CFR},n,m}\right)^*\right]}{\sum_{n^\mathrm{Rx}_\mathrm{A}=0}^{N^\mathrm{Rx}_\mathrm{A}-1}\left\lvert H^{n^\mathrm{Rx}_\mathrm{A}}_{\mathrm{CFR},n,m}\right\rvert^2}.
\end{equation}
In this equation, \mbox{$Y^{n^\mathrm{Rx}_\mathrm{A}}_{n,m}\in\mathbb{C}$} and \mbox{$H^{n^\mathrm{Rx}_\mathrm{A}}_{\mathrm{CFR},n,m}\in\mathbb{C}$} are the elements at the \mbox{$n\mathrm{th}$} row and \mbox{$m\mathrm{th}$} column of $\mathbf{Y}^{n^\mathrm{Rx}_\mathrm{A}}$ and $\mathbf{H}_\mathrm{CFR}^{n^\mathrm{Rx}_\mathrm{A}}$, respectively. The estimated transmit modulation symbols can be directly extracted from $\mathbf{Y}^\mathrm{MRC}$ to form an estimate of the transmit frame \mbox{$\mathbf{X}\in\mathbb{C}^{N\times M}$} that will be later also used for bistatic radar signal processing. If channel coding was applied, $\mathbf{Y}^\mathrm{MRC}$ must first undergo decoding to obtain the transmit bit stream estimate, which will then be re-encoded to form an estimate of $\mathbf{X}$ \cite{brunner2024}.

\subsubsection{Bistatic radar signal processing}\label{subsubsec:bistRadarSP}

Having estimated the transmit frame $\mathbf{X}$ as described in Section~\ref{subsubsec:commSP}, the bistatic signal processing starts by estimating the radar \ac{CFR} matrix \mbox{$\mathbf{D}^{n^\mathrm{Rx}_\mathrm{A}}\in\mathbb{C}^{N\times M}$} for every \mbox{$n^\mathrm{Rx}_\mathrm{A}\mathrm{th}$} receive channel. Since the \ac{ABE}-induced delay has already been considered during \ac{STO} correction as described in Section~\ref{subsubsec:sync}, only the magnitude of the \ac{ABE} \ac{CFR} must be calibrated to avoid distorting the ultimately obtained radar images. In this sense, the element at the \mbox{$n\mathrm{th}$} row and \mbox{$m\mathrm{th}$} column of $\mathbf{D}^{n^\mathrm{Rx}_\mathrm{A}}$ is calculated as
\begin{equation}\label{eq:D_radar}
	D^{n^\mathrm{Rx}_\mathrm{A}}_{n,m} = Y^{n^\mathrm{Rx}_\mathrm{A}}_{n,m}\left/\left(\left\lvert H^{n^\mathrm{Rx}_\mathrm{A}}_{\mathrm{ABE},n,m}\right\rvert X_{n,m}\right)\right.,
\end{equation}
where \mbox{$H^{n^\mathrm{Rx}_\mathrm{A}}_{\mathrm{ABE},n,m}\in\mathbb{C}$} is the element at the corresponding subcarrier and \ac{OFDM} symbol position of the \ac{ABE} \ac{CFR} matrix \mbox{$\mathbf{H}_\mathrm{ABE}^{n^\mathrm{Rx}_\mathrm{A}}\in\mathbb{C}^{N\times M}$}. The aforementioned \ac{ABE} \ac{CFR} matrix is associated with $h^{\mathrm{Rx},n^\mathrm{Rx}_\mathrm{A}}_\mathrm{ABE}(t)$ and can be estimated in advance via calibration measurements, e.g., over-the-air  \cite{vasanelli2020} or feeding a known broadband signal to all receive channels and estimating their response. Since the \ac{ABE} \ac{CFR} is time-invariant, $H^{n^\mathrm{Rx}_\mathrm{A}}_{\mathrm{ABE},n,m}$ only varies along with subcarrier index $n$, assuming the same complex values for different \ac{OFDM} symbol indexes $m$. It is also worth highlighting that only the absolute value of $H^{n^\mathrm{Rx}_\mathrm{A}}_{\mathrm{ABE},n,m}$ is taken to avoid distortion in the sidelobe profile of radar targets in the ultimately obtained range-Doppler shift radar images. This is done since the phase of the aforementioned \ac{CFR} is associated with delay experienced through convolution with the corresponding \ac{CIR}. Since this delay has already been compensated during synchronization, it does not have to be considered again when performing bistatic radar signal processing.

By performing the same bistatic radar signal processing steps as for the \ac{SISO} case on $\mathbf{D}^{n^\mathrm{Rx}_\mathrm{A}}$ \cite{giroto2023_EuMW,brunner2024}, a range-Doppler shift radar image \mbox{$\mathbf{I}^{n^\mathrm{Rx}_\mathrm{A}}\in\mathbb{C}^{N\times M}$} can be obtained for the \mbox{$n^\mathrm{Rx}_\mathrm{A}\mathrm{th}$} receive channel. In this radar image, the targets reflections associated with paths $p\in\{1,\dots,P-1\}$ will have image \ac{SNR} defined as
\begin{equation}\label{eq:imageSNR_rD}
	\mathit{SNR}_{\mathbf{I}^{n^\mathrm{Rx}_\mathrm{A}},p} = \frac{P_{\mathrm{Tx}}~(N^\mathrm{Tx}_\mathrm{A}~G_{\mathrm{Tx}})~G_{\mathrm{Rx}}~\sigma_{\mathrm{RCS},p}~\lambda_0^2~G_\mathrm{p}}{\left(4\pi\right)^3~{R^\mathrm{Tx}_p}^2~{R^\mathrm{Rx}_p}^2~k_\mathrm{B}~B~T_\mathrm{therm}~\mathit{NF}}
\end{equation}
In this equation, $G_{\mathrm{Tx}}$ and $G_{\mathrm{Rx}}$ are the single-element transmit and receive antenna gains, respectively. $\sigma_{\mathrm{RCS},p}$ is the \ac{RCS} of the $p\mathrm{th}$ target, which is assumed to be the same for every transmit and receive antenna pair for simplicity. \mbox{$G_\text{p}=NM$} is the range-Doppler shift radar processing gain. In addition, 
\begin{equation}
	R^\mathrm{Tx}_p = c_0~\tau^\mathrm{Tx}_p
\end{equation}
and
\begin{equation}
	R^\mathrm{Rx}_p = c_0~\tau^\mathrm{Rx}_p
\end{equation}
are the ranges from transmitter to the $p\mathrm{th}$ target and the $p\mathrm{th}$ target to the receiver, respectively. Similarly to \eqref{eq:delay_p}, they are defined assuming
\begin{equation}
	\tau^{\mathrm{Tx},n^\mathrm{Tx}_\mathrm{A}}_p \approx\tau^\mathrm{Tx}_p~\quad\forall~n^\mathrm{Tx}_\mathrm{A},p
\end{equation}
and
\begin{equation}
	\tau^{\mathrm{Rx},n^\mathrm{Rx}_\mathrm{A}}_p \approx\tau^\mathrm{Rx}_p~\quad\forall~n^\mathrm{Rx}_\mathrm{A},p.
\end{equation}
Finally, the term \mbox{$k_\text{B}~B~T_\text{therm}~\mathit{NF}$} in the denominator accounts for the \ac{AWGN} power. It is defined by the Boltzmann constant $k_\text{B}$, the previously defined \ac{OFDM} signal bandwidth $B$, the standard room temperature in Kelvin $T_\text{therm}$, and the overall receiver noise figure $\mathit{NF}$.

Next, Fourier beamforming \cite{nuss2017} is performed for \ac{DoA} estimation, which allows obtaining a three-dimensional radar image represented by the matrix \mbox{$\bm{\mathcal{I}}\in\mathbb{C}^{N\times M\times N^\mathrm{Rx}_\mathrm{S}}$}, where $N^\mathrm{Rx}_\mathrm{S}$ is the number of evaluated \acp{DoA}. Specifically, the depths of $\bm{\mathcal{I}}$ are associated with range, Doppler shift, and azimuth \ac{DoA}, respectively.

Assuming a steering vector $\mathbf{b}_{n^\mathrm{Rx}_\mathrm{S}}\in\mathbb{C}^{1\times N^\mathrm{Rx}_\mathrm{A}}$ for the \mbox{$n^\mathrm{Rx}_\mathrm{S}\mathrm{th}$} evaluated \ac{DoA} $\phi^\mathrm{Rx}_{n^\mathrm{Rx}_\mathrm{S}}$, whose \mbox{$n^\mathrm{Rx}_\mathrm{A}\mathrm{th}$} element is given by
\begin{equation}
	b^{\mathrm{Rx},n^\mathrm{Rx}_\mathrm{A}}_{n^\mathrm{Rx}_\mathrm{S}} = \e^{\im 2\pi\left(\rho^{n^\mathrm{Rx}_\mathrm{A}}_{x}/\lambda_0\right)\sin\left(\phi^\mathrm{Rx}_{n^\mathrm{Rx}_\mathrm{S}}\right)},
\end{equation}
the element at the \mbox{$n\mathrm{th}$} row, \mbox{$m\mathrm{th}$} column, and \mbox{$n^\mathrm{Rx}_\mathrm{S}\mathrm{th}$} depth of $\bm{\mathcal{I}}$ can be expressed as
\begin{align}\label{eq:I_rDa}
	\mathcal{I}_{n,m,n^\mathrm{Rx}_\mathrm{S}} &= \sum_{n^\mathrm{Rx}_\mathrm{S}=0}^{N^\mathrm{Rx}_\mathrm{S}-1}I^{n^\mathrm{Rx}_\mathrm{A}}_{n,m}~b^{\mathrm{Rx},n^\mathrm{Rx}_\mathrm{A}}_{n^\mathrm{Rx}_\mathrm{S}}\nonumber\\
	&= \sum_{n^\mathrm{Rx}_\mathrm{S}=0}^{N^\mathrm{Rx}_\mathrm{S}-1}I^{n^\mathrm{Rx}_\mathrm{A}}_{n,m}~\e^{\im 2\pi\left(\rho^{n^\mathrm{Rx}_\mathrm{A}}_{x}/\lambda_0\right)\sin\left(\phi^\mathrm{Rx}_{n^\mathrm{Rx}_\mathrm{S}}\right)}.
\end{align}
Extending the reasoning used for transmit beamforming to the Fourier beamforming-based \ac{DoA} estimation, the phase shifts $\e^{-\im 2\pi f_\mathrm{c}\tau^{\mathrm{Rx},n^\mathrm{Rx}_\mathrm{A}}_p}$ in \eqref{eq:ys} will lead to target peaks at for the same ranges and Doppler shifts as in the range-Doppler shift radar images $\mathbf{I}^{n^\mathrm{Rx}_\mathrm{A}}$, but now at every $n^\mathrm{Rx}_\mathrm{S}\mathrm{th}$ depth of $\bm{\mathcal{I}}$ that is associated with \ac{DoA} $\phi^\mathrm{Rx}_{n^\mathrm{Rx}_\mathrm{S}}$ such that \mbox{$\phi^\mathrm{Rx}_{n^\mathrm{Rx}_\mathrm{S}}=\phi^\mathrm{Rx}_{p}$}, where $\phi^\mathrm{Rx}_{p}$ is the actual \ac{DoA} w.r.t. the receiver associated with the $p\mathrm{th}$ path. Target reflections in this radar image associated with paths $p\in\{1,\dots,P-1\}$ have image \ac{SNR} given by
\begin{equation}\label{eq:imageSNR_rDA}
	\mathit{SNR}_{\bm{\mathcal{I}},p} = \frac{P_{\mathrm{Tx}}~(N^\mathrm{Tx}_\mathrm{A}~G_{\mathrm{Tx}})~(N^\mathrm{Rx}_\mathrm{A}~G_{\mathrm{Rx}})~\sigma_{\mathrm{RCS},p}~\lambda_0^2~G_\mathrm{p}}{\left(4\pi\right)^3~{R^\mathrm{Tx}_p}^2~{R^\mathrm{Rx}_p}^2~k_\mathrm{B}~B~T_\mathrm{therm}~\mathit{NF}},
\end{equation}
where a gain of factor $N^\mathrm{Rx}_\mathrm{A}$ w.r.t. \eqref{eq:imageSNR_rD} is observed due to the coherent processing of $N^\mathrm{Rx}_\mathrm{A}$ range-Doppler shift radar images $\mathbf{I}^{n^\mathrm{Rx}_\mathrm{A}}$ for \ac{DoA} estimation.

\begin{table}[!t]
	\renewcommand{\arraystretch}{1.5}
	\arrayrulecolor[HTML]{708090}
	\setlength{\arrayrulewidth}{.1mm}
	\setlength{\tabcolsep}{4pt}
	
	\centering
	\captionsetup{width=43pc,justification=centering,labelsep=newline}
	\caption{\textsc{OFDM Signal and ISAC Performance Parameters}}
	\label{tab:ofdmParameters}
		\begin{tabular}{|cc||cc|}
			\hhline{|====|}
			\multicolumn{2}{|c||}{\textbf{OFDM signal parameters}} & \multicolumn{2}{c|}{\textbf{ISAC performance parameters}} \\
			\hhline{|====|}
			\multicolumn{1}{|c|}{$f_\text{c}$} & $\SI{27.5}{\giga\hertz}$ & \multicolumn{1}{|c|}{$\mathcal{R}_\mathrm{comm}$} & $\SI{0.39}{Gbit/s}$ \\
			\hline
			\multicolumn{1}{|c|}{$B$} & $\SI{491.52}{\mega\hertz}$ & \multicolumn{1}{|c|}{$G_\mathrm{p}$} & $\SI{60.22}{dB}$ \\
			\hline
			\multicolumn{1}{|c|}{$N$} & $2048$ & \multicolumn{1}{|c|}{$\Delta R$} & $\SI{0.61}{\meter}$ \\
			\hline
			\multicolumn{1}{|c|}{$\Delta f$} & $\SI{240}{\kilo\hertz}$ & \multicolumn{1}{|c|}{$R_\mathrm{max,ua}$} & $\SI{1249.14}{\meter}$ \\
			\hline
			\multicolumn{1}{|c|}{$N_\mathrm{CP}$} & $512$ & \multicolumn{1}{|c|}{$R_\mathrm{max,ISI}$} & $\SI{312.28}{\meter}$ \\
			\hline
			\multicolumn{1}{|c|}{$M$} & $512$ & \multicolumn{1}{|c|}{$\Delta f_\mathrm{D}$} & $\SI{375}{\hertz}$ \\
			\hline
			\multicolumn{1}{|c|}{$\Delta N_\mathrm{pil}$} & $2$ & \multicolumn{1}{|c|}{$f_\mathrm{D,max,ua}$} & $\pm\SI{96}{\kilo\hertz}$ \\
			\hline
			\multicolumn{1}{|c|}{$\Delta M_\mathrm{pil}$} & $2$ & \multicolumn{1}{|c|}{$f_\mathrm{D,max,ICI}$} & $\pm\SI{24}{\kilo\hertz}$ \\
			\hline
			\multicolumn{1}{|c|}{$N_\mathrm{pil}$} & $1024$ & \multicolumn{1}{|c|}{$\Delta \phi^\mathrm{Rx}$} & \ang{14.32} \\
			\hline
			\multicolumn{1}{|c|}{$M_\mathrm{pil}$} & $256$ & \multicolumn{1}{|c|}{$\phi^\mathrm{Rx}_\mathrm{max,ua}$} & $\pm$\ang{90} \\
			\hhline{|====|}      
		\end{tabular}
\end{table}

\section{Numerical analysis}\label{sec:simResults}

To analyze the performance of the proposed bistatic \ac{MIMO}-\ac{OFDM}-based \ac{ISAC} system concept, \mbox{$N^\mathrm{Tx}_\mathrm{A}=4$} transmit and \mbox{$N^\mathrm{Rx}_\mathrm{A}=8$} receive channels are considered, with both transmit and receive arrays assumed to be \acp{ULA} with a \mbox{$\lambda_0/2$} element spacing. Furthermore, it is assumed that \ac{BP} sampling at a digital \ac{IF} of \mbox{$f_\mathrm{IF}=\SI{3.68}{\giga\hertz}$} is performed. This specific digital \ac{IF} is used in the described testbed in \cite{nuss2024} that is later used for measurement-based validation in Section~\ref{sec:measResults}.

The adopted \ac{OFDM} signal parameters including the ones discussed so far, besides subcarrier spacing \mbox{$\Delta f=B/N$} and \ac{CP} length \mbox{$N_\mathrm{CP}\in\mathbb{N}_{\geq 0}$}, are listed in Table~\ref{tab:ofdmParameters}. In addition, \ac{QPSK} modulation for all subcarriers and an \ac{LDPC} channel code \cite{miao2024} of rate $2/3$ and the parity check matrix for the case with $64800$ bits in an \ac{LDPC} code block from \cite{ETSI302307} were assumed, which resulted in the \ac{ISAC} performance parameters calculated according to \cite{giroto2023_EuMW,vasanelli2020} and listed in the same table. These are namely communication data rate at 100\% duty cycle, $\mathcal{R}_\mathrm{comm}$, processing gain $G_\mathrm{p}$ considering the \ac{DoA} estimation gain, range resolution $\Delta R$, maximum unambiguous range $R_\mathrm{max,ua}$, maximum \ac{ISI}-free range $R_\mathrm{max,ISI}$, Doppler shift resolution $\Delta f_\mathrm{D}$, maximum unambiguous Doppler shift $f_\mathrm{D,max,ua}$, maximum \ac{ICI}-free Doppler shift $f_\mathrm{D,max,ICI}$, Rx azimuth resolution $\Delta \phi^\mathrm{Rx}$, and maximum unambiguous Rx azimuth $\phi^\mathrm{Rx}_\mathrm{max,ua}$.

Most relevant aspects of bistatic \ac{OFDM}-based \ac{ISAC} system performance have been analyzed in \cite{brunner2024} for the \ac{SISO} case. In the adopted \ac{MIMO} architecture, the main difference is  the \ac{MRC} diversity gain obtained according to Section~\ref{subsubsec:commSP} for communication, the \ac{DoA} estimation capability for radar as described in Section~\ref{subsubsec:bistRadarSP}, besides the distributed synchronization discussed in Section~\ref{subsubsec:sync}. In this article, the focus is placed on the latter. Since all \mbox{$N^\mathrm{Tx}_\mathrm{A}=4$} transmit channels and \mbox{$N^\mathrm{Rx}_\mathrm{A}=8$} receive channels share the same \ac{LO} reference and sampling clocks locally at the transmitter and receiver, respectively, the same \ac{CFO} and \ac{SFO} will be experienced at all receive channels as discussed in Section~\ref{subsubsec:sync}. Therefore, no further analysis of these impairments is performed in this article and the reader is referred to \cite{brunner2024,giroto2024_tmtt}. An additional hardware impairment that may impair the performance of the considered bistatic \ac{MIMO}-\ac{OFDM}-based \ac{ISAC} system is \ac{PN}. The effect of \ac{PN} on bistatic \ac{OFDM}-based \ac{ISAC} has been investigated assuming \ac{PLL}-based oscillators in a \ac{SISO} architecture in \cite{giroto2024PN}, showing that it leads to \ac{ICI} and \ac{CPE}, which degrade the performance of range and Doppler shift estimations, respectively. In the \ac{MIMO} case considered in this article, \ac{PN} may additionally degrade both the transmit beamforming and the \ac{DoA} estimation at the receiver side. The extent of the degradation will depend on whether independent \acp{LO} are adopted for each receive channel or whether a single \ac{LO} is distributed to the mixers of all $N^\mathrm{Rx}_\mathrm{A}$ receive channels \cite{puglielli2016}. In the first case, the uncorrelated \ac{PN} of the different channels tends to average out at the beamforming or \ac{DoA} directions \cite{hoehne2010}. As for the latter case, although the different delays $\tau_\mathrm{ABE}^{n^\mathrm{Rx}_\mathrm{A}}$ experienced at each receive channel lead to reduction of the correlation of their \ac{PN} contributions, performance degradation may still occur if a low-\ac{PN} \ac{LO} source is not adopted \cite{collmann2025}. This, however, is left as an open topic for a future study since a single, low-\ac{PN} \ac{LO} source was adopted for the measurements later discussed in Section~\ref{sec:measResults}.

\begin{figure}[!t]
	\centering
	
	\subfloat[ ]{
		
		\psfrag{-6}[][]{\scriptsize -$6$}
		\psfrag{-3}[][]{\scriptsize -$3$}
		\psfrag{0}[][]{\scriptsize $0$}
		\psfrag{3}[][]{\scriptsize $3$}
		
		\psfrag{-160}[][]{\scriptsize -$160$}
		\psfrag{-128}[][]{\scriptsize -$128$}
		\psfrag{-96}[][]{\scriptsize -$96$}
		\psfrag{-64}[][]{\scriptsize -$64$}
		\psfrag{-32}[][]{\scriptsize -$32$}
		\psfrag{0}[][]{\scriptsize $0$}
		
		\psfrag{log10(S/Ts)}[][]{\footnotesize $\log_{10}\left(\sigma_\tau/T_\mathrm{s}\right)$}
		\psfrag{EVM (dB)}[][]{\footnotesize EVM (dB)}	
		
		\includegraphics[height=3.75cm]{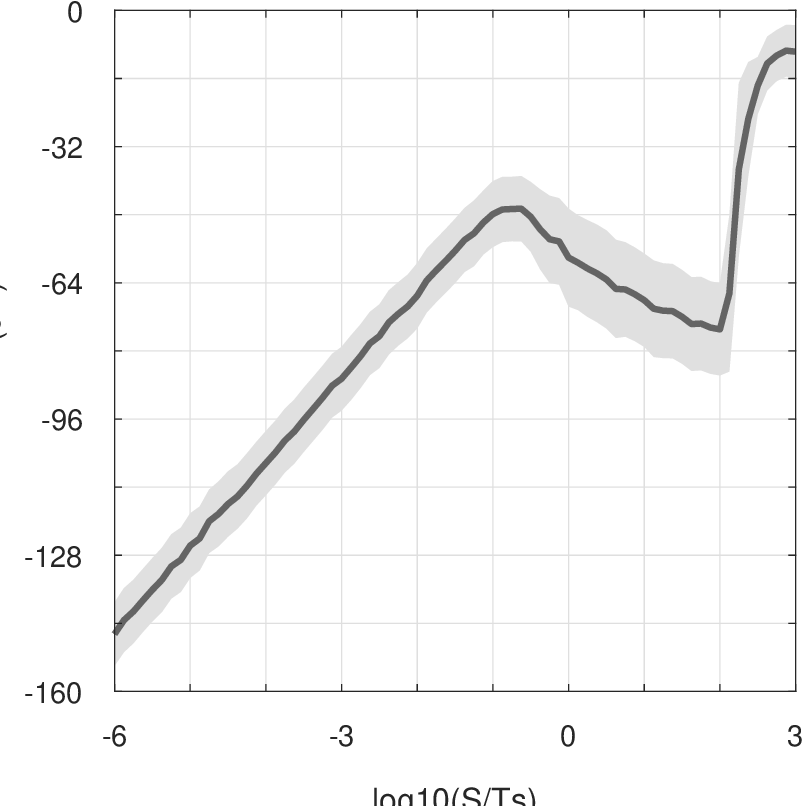}\label{fig:EVM}
		
	}\hspace{0.1cm}
	\subfloat[ ]{
		
		\psfrag{-6}[][]{\scriptsize -$6$}
		\psfrag{-3}[][]{\scriptsize -$3$}
		\psfrag{0}[][]{\scriptsize $0$}
		\psfrag{3}[][]{\scriptsize $3$}
		
		\psfrag{0}[][]{\scriptsize $0$}
		\psfrag{0.02}[][]{\scriptsize $0.02$}
		\psfrag{0.04}[][]{\scriptsize $0.04$}
		\psfrag{0.06}[][]{\scriptsize $0.06$}
		
		\psfrag{log10(S/Ts)}[][]{\footnotesize $\log_{10}\left(\sigma_\tau/T_\mathrm{s}\right)$}
		\psfrag{BER}[][]{\footnotesize BER}	
		
		\includegraphics[height=3.725cm]{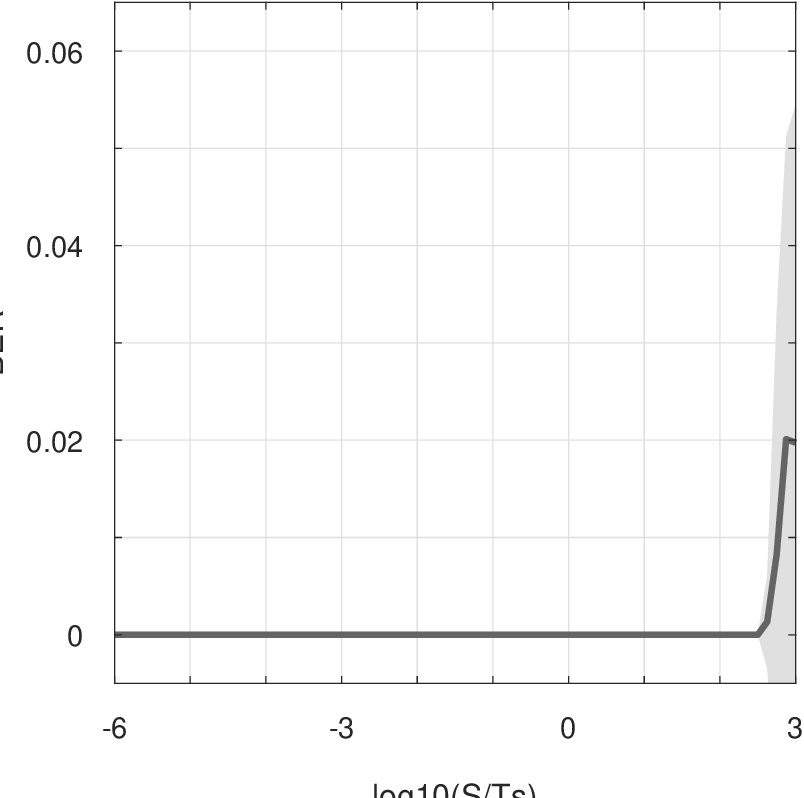}\label{fig:BER}	
		
	}		
	
	\captionsetup{justification=raggedright,labelsep=period,singlelinecheck=false}
	\caption{\ EVM (a) and BER (b) as functions of the normalized delay standard deviation $\sigma_\tau$ by the sampling period $T_\mathrm{s}$. Both parameters were calculated for an ideal, noiseless channel. The continuous lines represent the mean value of the calculated parameters and the shading in the background represents the standard deviation.}\label{fig:EVM_BER}
	
\end{figure}

Regarding \ac{STO}, however, an analysis of the robustness of the proposed to mismatches among the delays $\tau_\mathrm{ABE}^{n^\mathrm{Rx}_\mathrm{A}}$ associated with the $N^\mathrm{Rx}_\mathrm{A}$ receive channels of the \ac{ABE} becomes necessary as they may not be perfectly aligned even after synchronization as described in Section~\ref{subsubsec:sync}. For that purpose, it is henceforth assumed that \ac{CFO} and \ac{SFO} are absent and that perfect time synchronization is performed for the receive channel \mbox{$n^\mathrm{Rx}_\mathrm{A}=0$}. For all other \mbox{$N^\mathrm{Rx}_\mathrm{A}=7$} channels, i.e., \mbox{$n^\mathrm{Rx}_\mathrm{A}\in\{1,\dots,7\}$}, random delays defined according to a Rayleigh distribution with standard deviation $\sigma_\tau$ considered. Due to the use of a digital \ac{IF}, a corresponding angular standard deviation $\sigma_\theta$ will be experienced among the channels. The choice of the Rayleigh distribution is made to ensure that no \ac{ISI} immediately occurs for receive channels having smaller delay than \mbox{$n^\mathrm{Rx}_\mathrm{A}=0$}. Based on the previous assumptions, the communication and radar sensing performances of the proposed bistatic \ac{MIMO}-\ac{OFDM}-based \ac{ISAC} system are analyzed in Sections~\ref{subsec:simResults_comm} and \ref{subsec:simResults_rad}. In these analysis, the critical sampling period \mbox{$T_\mathrm{s}=1/B$} is used as a reference to which the delay standard deviation is compared.

\begin{figure*}[!t]
	\centering
	\subfloat[ ]{
		
		\psfrag{-6}[][]{\scriptsize -$6$}
		\psfrag{-3}[][]{\scriptsize -$3$}
		\psfrag{0}[][]{\scriptsize $0$}
		\psfrag{3}[][]{\scriptsize $3$}
		
		\psfrag{-25}[][]{\scriptsize -$25$}
		\psfrag{-20}[][]{\scriptsize -$20$}
		\psfrag{-15}[][]{\scriptsize -$15$}
		\psfrag{-10}[][]{\scriptsize -$10$}
		\psfrag{-5}[][]{\scriptsize -$5$}
		\psfrag{0}[][]{\scriptsize $0$}
		
		\psfrag{log10(S/Ts)}[][]{\footnotesize $\log_{10}\left(\sigma_\tau/T_\mathrm{s}\right)$}
		\psfrag{PPLR (dB)}[][]{\footnotesize $\mathrm{PPLR~(dB)}$}			
		
		\includegraphics[width=3.25cm]{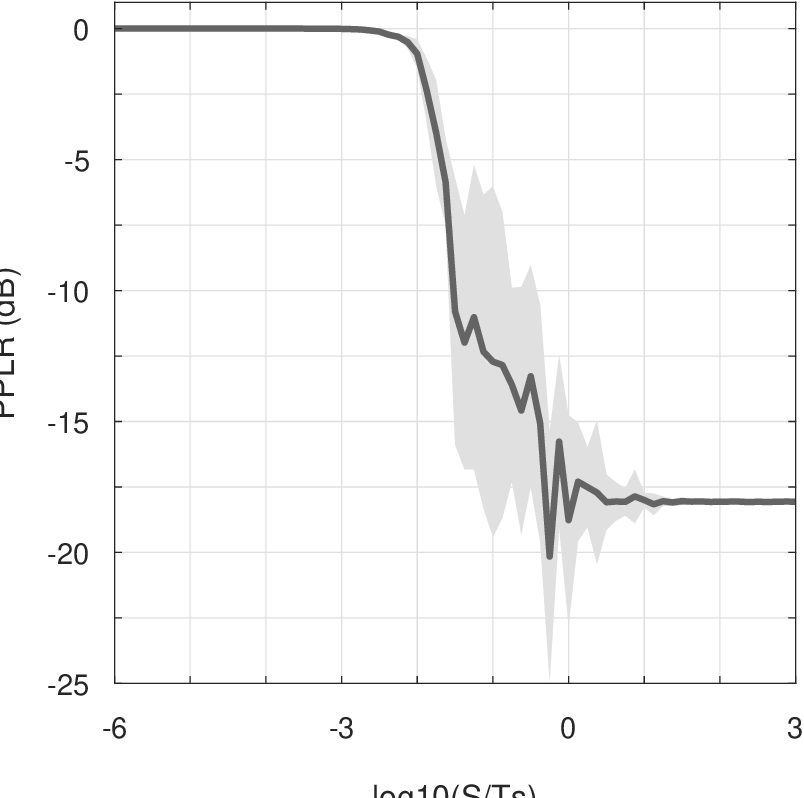}			
	}\hfill
	\subfloat[ ]{
		
		\psfrag{-6}[][]{\scriptsize -$6$}
		\psfrag{-3}[][]{\scriptsize -$3$}
		\psfrag{0}[][]{\scriptsize $0$}
		\psfrag{3}[][]{\scriptsize $3$}
		
		\psfrag{-20}[][]{\scriptsize -$20$}
		\psfrag{-14}[][]{\scriptsize -$14$}
		\psfrag{-8}[][]{\scriptsize -$8$}
		\psfrag{-2}[][]{\scriptsize -$2$}
		\psfrag{4}[][]{\scriptsize $4$}
		\psfrag{10}[][]{\scriptsize $10$}
		
		\psfrag{log10(S/Ts)}[][]{\footnotesize $\log_{10}\left(\sigma_\tau/T_\mathrm{s}\right)$}
		\psfrag{Range PSLR (dB)}[][]{\footnotesize Range $\mathrm{PSLR~(dB)}$}	
		
		\includegraphics[width=3.25cm]{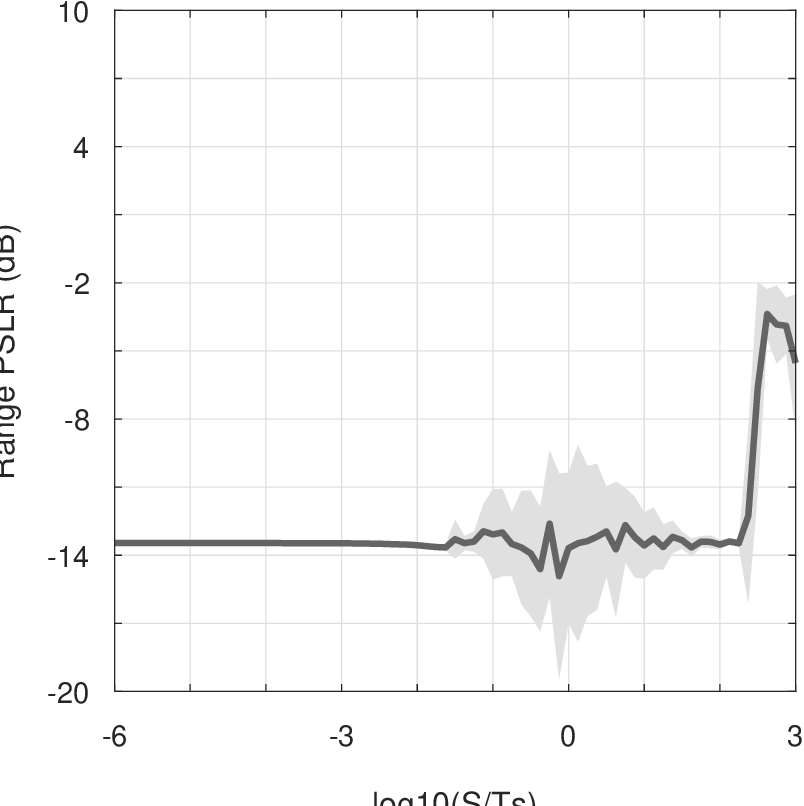}	
		
	}\hfill
	\subfloat[ ]{
		
		\psfrag{-6}[][]{\scriptsize -$6$}
		\psfrag{-3}[][]{\scriptsize -$3$}
		\psfrag{0}[][]{\scriptsize $0$}
		\psfrag{3}[][]{\scriptsize $3$}
		
		\psfrag{-15}[][]{\scriptsize -$15$}
		\psfrag{-9}[][]{\scriptsize -$9$}
		\psfrag{-3}[][]{\scriptsize -$3$}
		\psfrag{3}[][]{\scriptsize $3$}
		\psfrag{9}[][]{\scriptsize $9$}
		\psfrag{15}[][]{\scriptsize $15$}
		
		\psfrag{log10(S/Ts)}[][]{\footnotesize $\log_{10}\left(\sigma_\tau/T_\mathrm{s}\right)$}
		\psfrag{Range ISLR (dB)}[][]{\footnotesize Range $\mathrm{ISLR~(dB)}$}	
		
		\includegraphics[width=3.25cm]{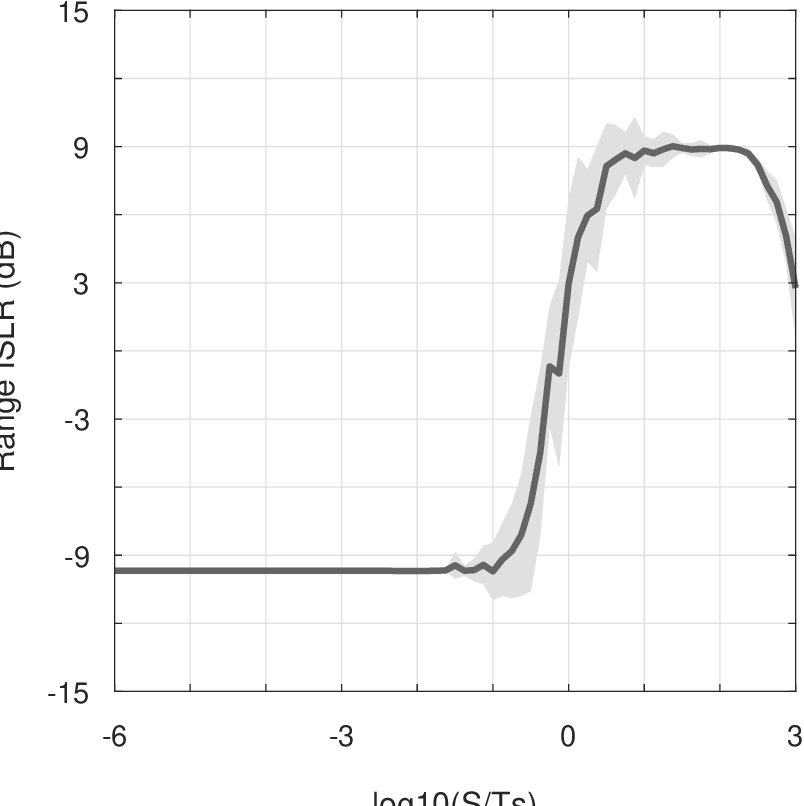}
		
	}\hfill
	\subfloat[ ]{
		
		\psfrag{-6}[][]{\scriptsize -$6$}
		\psfrag{-3}[][]{\scriptsize -$3$}
		\psfrag{0}[][]{\scriptsize $0$}
		\psfrag{3}[][]{\scriptsize $3$}
		
		\psfrag{-20}[][]{\scriptsize -$20$}
		\psfrag{-14}[][]{\scriptsize -$14$}
		\psfrag{-8}[][]{\scriptsize -$8$}
		\psfrag{-2}[][]{\scriptsize -$2$}
		\psfrag{4}[][]{\scriptsize $4$}
		\psfrag{10}[][]{\scriptsize $10$}
		
		\psfrag{log10(S/Ts)}[][]{\footnotesize $\log_{10}\left(\sigma_\tau/T_\mathrm{s}\right)$}
		\psfrag{Azimuth PSLR (dB)}[][]{\footnotesize Azimuth $\mathrm{PSLR~(dB)}$}	
		
		\includegraphics[width=3.25cm]{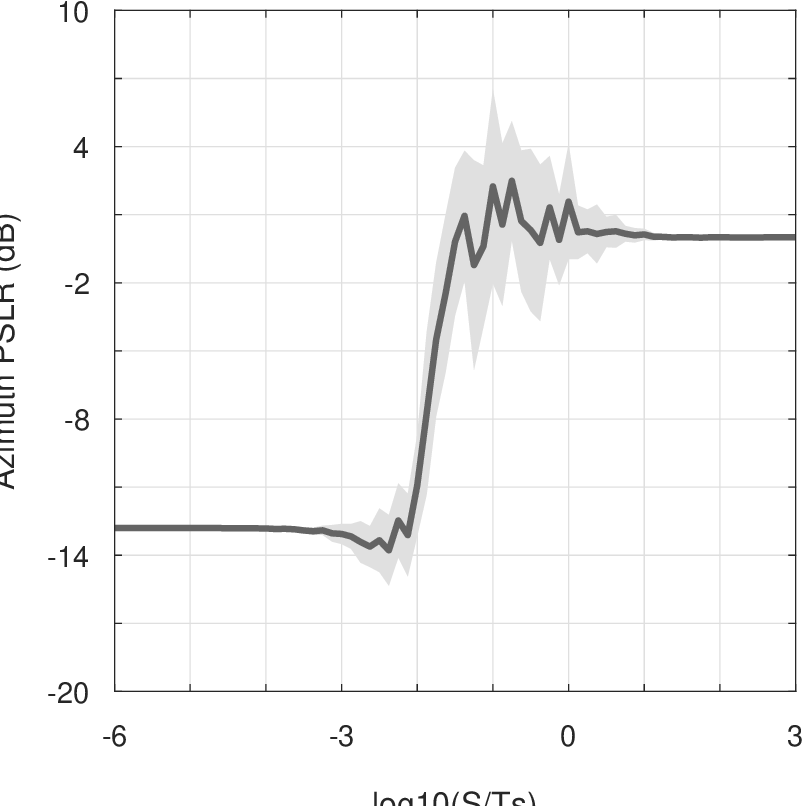}	
		
	}\hfill
	\subfloat[ ]{
		
		\psfrag{-6}[][]{\scriptsize -$6$}
		\psfrag{-3}[][]{\scriptsize -$3$}
		\psfrag{0}[][]{\scriptsize $0$}
		\psfrag{3}[][]{\scriptsize $3$}
		
		\psfrag{-15}[][]{\scriptsize -$15$}
		\psfrag{-9}[][]{\scriptsize -$9$}
		\psfrag{-3}[][]{\scriptsize -$3$}
		\psfrag{3}[][]{\scriptsize $3$}
		\psfrag{9}[][]{\scriptsize $9$}
		\psfrag{15}[][]{\scriptsize $15$}
		
		\psfrag{log10(S/Ts)}[][]{\footnotesize $\log_{10}\left(\sigma_\tau/T_\mathrm{s}\right)$}
		\psfrag{Azimuth ISLR (dB)}[][]{\footnotesize Azimuth $\mathrm{ISLR~(dB)}$}
		
		\includegraphics[width=3.25cm]{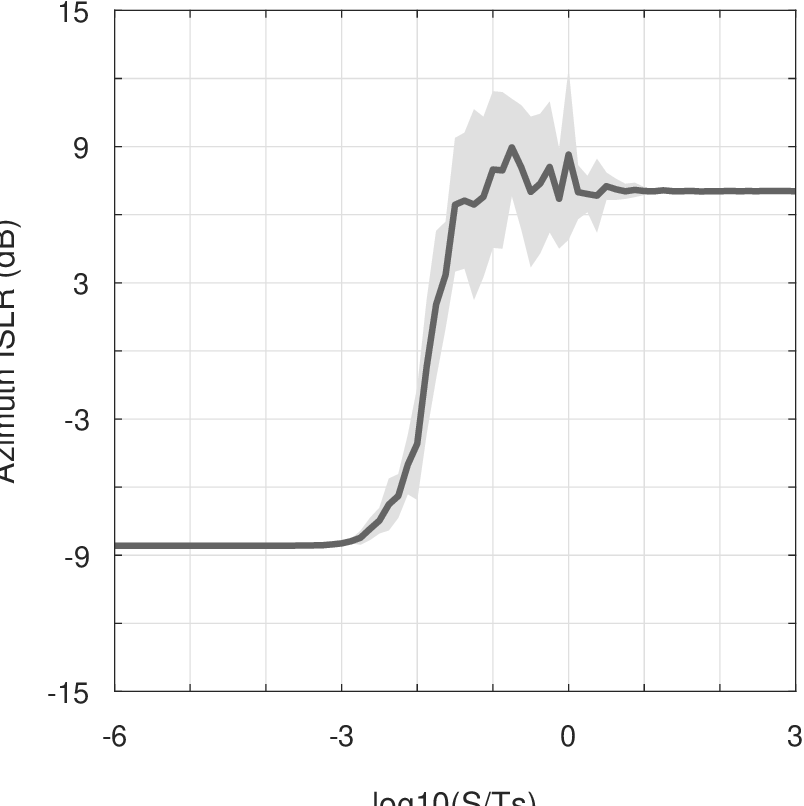}		
		
	}
	
	\captionsetup{justification=raggedright,labelsep=period,singlelinecheck=false}
	\caption{\ PPLR (a), range PSLR (b) and ISLR (c), and Doppler shift PSLR (e) and ISLR (d) as functions of the normalized delay standard deviation $\sigma_\tau$ by the sampling period $T_\mathrm{s}$. The continuous lines represent the mean value of the calculated parameters and the shading in the background represents the standard deviation.}\label{fig:ML_SL}
	
\end{figure*}

\subsubsection{Communication performance}\label{subsec:simResults_comm}

Fig.~\ref{fig:EVM_BER} shows the \ac{EVM} and \ac{BER} simulated as functions of the normalized delay standard deviation by the sampling period, i.e., $\sigma_\tau/T_\mathrm{s}$. The obtained results assume \ac{MRC} among the \mbox{$N^\mathrm{Rx}_\mathrm{A}=8$} receive channels according to \eqref{eq:MRC}. Fig.~\ref{fig:EVM} shows that the \ac{EVM} increases linearly until around \mbox{$\sigma_\tau=10^{-1}~T_\mathrm{s}$}. This corresponds to $10^{-1}$ discrete samples or $\SI{203.45}{\pico\second}$ and is associated with a phase standard deviation of \mbox{$\sigma_\theta=\ang{139.03}$} at the digital \ac{IF} of \mbox{$f_\mathrm{IF}=\SI{3.68}{\giga\hertz}$}. Although a large deviation between the phases of the channels is experienced, this delay standard deviation does not result in \ac{ISI} and therefore allows the \ac{MRC} processing to perform sufficiently well, which is supported by the low \ac{EVM} values that are equal to or lower than $\SI{-47.84}{dB}$ in the aforementioned region. Between \mbox{$\sigma_\tau=10^{-1}~T_\mathrm{s}$} and \mbox{$10^{2}~T_\mathrm{s}$}, the \ac{EVM} trend changes and its values decrease until as low as $\SI{-74.94}{dB}$. For \mbox{$\sigma_\tau>10^{2}~T_\mathrm{s}$}, the \ac{EVM} starts increasing again as such high delay standard deviations may result in \ac{ISI} for some of the receive channels. Although distortions of the \ac{QPSK} constellation were observed, it can be seen in Fig.~\ref{fig:BER} that the adopted \ac{LDPC} channel code can ensure null \ac{BER} until around \mbox{$\sigma_\tau=10^{2.5}~T_\mathrm{s}$}. Afterwards the experienced \ac{ISI} at some of the receive channels can no longer be handled by the channel code and the \ac{BER} increases up to $0.02$ at \mbox{$\sigma_\tau=10^{3}~T_\mathrm{s}$}.

\begin{figure}[!t]
	\centering
	
	\psfrag{-6}[][]{\scriptsize -$6$}
	\psfrag{-3}[][]{\scriptsize -$3$}
	\psfrag{0}[][]{\scriptsize $0$}
	\psfrag{3}[][]{\scriptsize $3$}
	
	\psfrag{5}[][]{\scriptsize $5$}
	\psfrag{24}[][]{\scriptsize $24$}
	\psfrag{43}[][]{\scriptsize $43$}
	\psfrag{62}[][]{\scriptsize $62$}
	\psfrag{81}[][]{\scriptsize $81$}
	\psfrag{100}[][]{\scriptsize $100$}
	
	\psfrag{log10(S/Ts)}[][]{\footnotesize $\log_{10}\left(\sigma_\tau/T_\mathrm{s}\right)$}
	\psfrag{Mean image SIR (dB)}[][]{\footnotesize Mean image SIR (dB)}	
	
	\includegraphics[height=4.75cm]{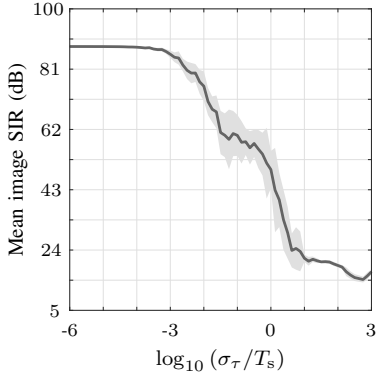}
	
	\captionsetup{justification=raggedright,labelsep=period,singlelinecheck=false}
	\caption{\ Mean image SIR as a function of the normalized delay standard deviation $\sigma_\tau$ by the sampling period $T_\mathrm{s}$. The continuous lines represent the mean value of the calculated parameters and the shading in the background represents the standard deviation.}\label{fig:imageSIR_mean}
	
\end{figure}

\subsubsection{Radar sensing performance}\label{subsec:simResults_rad}

To solely analyze the influence of \ac{STO} mismatch among the receive channels, \textit{genie-aided} decoding is assumed to allow perfect reconstruction of the transmit frame needed for bistatic radar signal processing \cite{giroto2023_EuMW,brunner2024}. In addition, a single, static radar target is considered and \ac{AWGN} is disregarded to allow solely focusing on the \ac{STO} effect. Fig.~\ref{fig:ML_SL} shows the \ac{PPLR}, range \ac{PSLR} and \ac{ISLR}, as well as azimuth \ac{PSLR} and \ac{ISLR} calculated for the considered target in a range-azimuth cut of the radar image calculated as in \eqref{eq:I_rDa} at zero Doppler shift as a function of the normalized delay standard deviation among receive channels \mbox{$n^\mathrm{Rx}_\mathrm{A}\in\{1,\dots,7\}$} by the sampling period $T_\mathrm{s}$. For the calculation of these parameters, rectangular windowing was used for both range and azimuth processing to avoid window-specific sidelobe supression. The obtained results show that, until around \mbox{$\sigma_\theta=10^{-3}~T_\mathrm{s}$}, which corresponds to $10^{-3}$ discrete samples or $\SI{20.35}{\pico\second}$ and is associated with a phase standard deviation of \mbox{$\sigma_\theta=\ang{2.70}$} at the digital \ac{IF} of \mbox{$f_\mathrm{IF}=\SI{3.68}{\giga\hertz}$}, no significant mainlobe or sidelobe degradation is observed. Afterwards, azimuth \ac{PSLR} and \ac{ISLR} are the first to degrade, with visible changes in sidelobes observable before \mbox{$\sigma_\tau=10^{-2}~T_\mathrm{s}$}, which is associated with $\sigma_\theta=\ang{26.96}$. After this point, \ac{PPLR} degradation is no longer negligible, with $\SI{3}{dB}$ peak power loss being reached at a delay standard deviation of around \mbox{$\sigma_\tau=6\cdot10^{-2}~T_\mathrm{s}$} and \mbox{$\sigma_\theta=\ang{100}$}. Both range \ac{PSLR} and \ac{ISLR} only start to degrade at the same aforementioned $\sigma_\tau$ value. However, while the \ac{ISLR} continuously increases, which shows that the overall sidelobe level becomes higher, the range \ac{PSLR} oscillates and only starts to continuously increase at \mbox{$\sigma_\tau=177.83~T_\mathrm{s}$}, which corresponds to \mbox{$\sigma_\theta=\ang{219.13}$}. At this high delay standard deviation, \ac{ISI} may not be entirely ruled out for all receive channels as \ac{CP} of length  \mbox{$N_\mathrm{CP}=512$} was adopted.

Combined with the earlier degradation of azimuth \ac{PSLR} and \ac{ISLR}, the \ac{PPLR} and range \ac{PSLR} and \ac{ISLR} results show that the azimuth estimation is more severely affected by the mismatch of the experienced \acp{STO} among the \mbox{$N^\mathrm{Rx}_\mathrm{A}=8$} receive channels due to the experienced phase mismatch among the receive channels. While the achieved results seem to indicate that that tolerable performance degradation is attained even for high phase rotations, as is the case for \mbox{$\sigma_\tau=6\cdot10^{-2}~T_\mathrm{s}$} and \mbox{$\sigma_\theta=\ang{100}$}, one must consider that the use of rectangular windowing results in high sinc-shaped sidelobes that may mask the overall radar sensing performance degradation. In this sense, a further performance parameter, namely the image \ac{SIR} is analyzed as follows.

Fig.~\ref{fig:imageSIR_mean} shows the mean image \ac{SIR} as a function of $\sigma_\tau/T_\mathrm{s}$. The results were obtained from simulations with the considered single, static target and assuming the use of Chebyshev windowing with $\SI{100}{dB}$ sidelobe suppression for both range and azimuth processing to ensure that only distortion in the cut of the radar image at a Doppler shift of $\SI{0}{\hertz}$, and not range or azimuth sidelobes are considered. The mean image \ac{SIR} is calculated as the ratio between the peak power of the target at $\SI{0}{\meter}$ range and $\ang{0}$ azimuth, which is associated with the receive channel \mbox{$n^\mathrm{Rx}_\mathrm{A}=0$} that has unbiased synchronization, and the average power at the rest of the radar image cut. The obtained results show that only negligible mean image \ac{SIR} degradation is observed for \mbox{$\sigma_\tau\leq10^{-3.25}~T_\mathrm{s}$}, which corresponds to \mbox{$\sigma_\theta\leq\ang{1.52}$}. Since the aforementioned values are low, they do not lead to either significant range offsets nor phase rotation among the receive channels, which explains the nearly constant mean image \ac{SIR}. For \mbox{$\sigma_\tau>10^{-2}~T_\mathrm{s}$}, the mean image \ac{SIR} starts to decrease rapidly, with a $\SI{10}{dB}$ reduction w.r.t. to the highest achievable mean image \ac{SIR} being already experienced at \mbox{$\sigma_\tau\leq10^{-2}~T_\mathrm{s}$}. Since, however, the overall performance degradation in the aforementioned region is moderate and mainly due to phase mismatches among the receive channels, it can be considered that \ac{STO} can be sufficiently compensated afterwards at the pilot-based fine tuning discussed in Section~\ref{subsubsec:sync} since it may achieve sub-sample accuracy, e.g., by using \ac{CZT} as discussed in \cite{giroto2024_tmtt}.

\begin{figure}[!t]
	\centering

	\includegraphics[width=8.75cm]{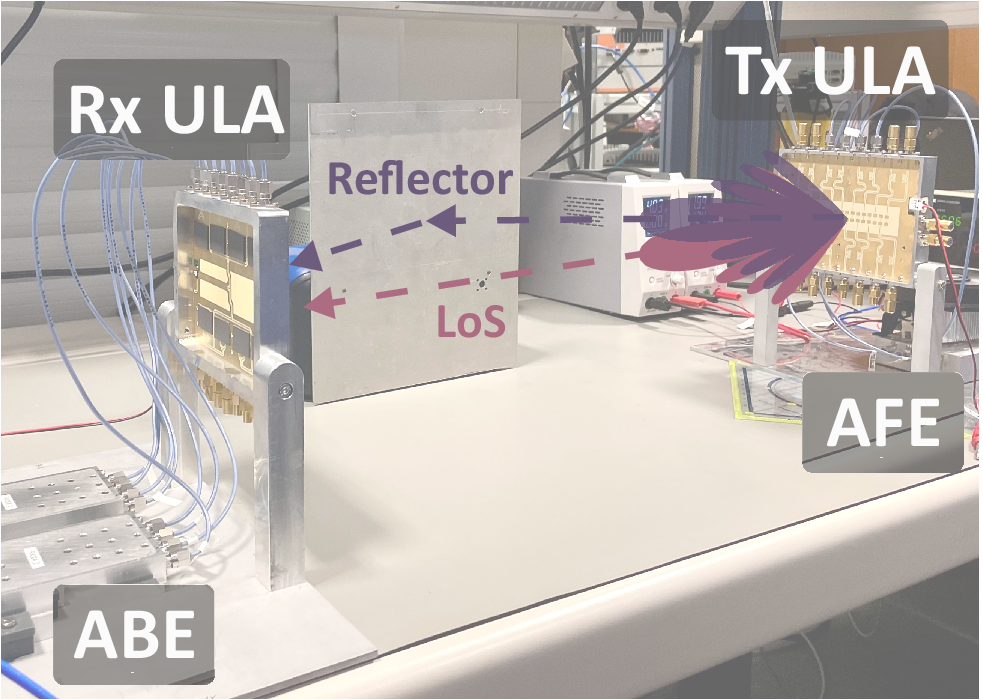}

	\captionsetup{justification=raggedright,labelsep=period,singlelinecheck=false}
	\caption{\ Bistatic MIMO-OFDM-based ISAC measurement setup.}\label{fig:measSetupSmall}
\end{figure}

\section{Measurement-based verification}\label{sec:measResults}

To validate the proposed bistatic \ac{MIMO}-\ac{OFDM}-based \ac{ISAC} system concept, the setup shown in Fig.~\ref{fig:measSetupSmall} was adopted. It has the same considered transmit and receive arrays with \mbox{$N^\mathrm{Tx}_\mathrm{A}=4$} and $N^\mathrm{Rx}_\mathrm{A}=8$ antenna elements, respectively, as assumed in Section~\ref{sec:simResults}. All aforementioned elements are patch antennas that were selected from \acp{ULA} designed with an element spacing of \mbox{$\lambda_0/2$} at $\SI{28}{\giga\hertz}$. Further details on the adopted \acp{ULA} and their associated \acp{AFE} and \acp{ABE} can be found in \cite{eisenbeis2021_diss}. For signal generation and sampling at the receiver side, the described testbed in \cite{nuss2024}, with specific components further analyzed in \cite{karle2024_1,karle2024_2,scheder2024}, was adopted. As in Section~\ref{sec:simResults}, \ac{BP} sampling with a digital \ac{IF} of \mbox{$f_\mathrm{IF}=\SI{3.68}{\giga\hertz}$} was performed, and the same \ac{OFDM} signal parameters from Table~\ref{tab:ofdmParameters} were adopted. In the adopted measurement scenario, the transmit \ac{ULA} was placed at approximately $\SI{55}{\centi\meter}$ and an approximate \ac{DoA} of $\ang{3}$ w.r.t. the receive \ac{ULA}. In addition, a reflector with an approximate bistatic range of $\SI{70}{\centi\meter}$ and an approximate \ac{DoA} of $\ang{-20}$ was used as a static radar target. At the transmitter side, the \acp{AFE} associated with each transmit channel were calibrated and beamforming was performed towards the \acp{DoD} of $\ang{0}$ and $\ang{30}$, which approximately correspond to the \acp{DoD} of the \ac{LoS} reference path towards the receiver and the reflector, respectively.

\begin{figure}[!t]
	\centering
	
	\psfrag{-12}[c][c]{\footnotesize -$12$}
	\psfrag{-9}[c][c]{\footnotesize -$9$}
	\psfrag{-6}[c][c]{\footnotesize -$6$}
	\psfrag{-3}[c][c]{\footnotesize -$3$}
	\psfrag{0}[c][c]{\footnotesize $0$}
	
	\psfrag{-200}[c][c]{\footnotesize -$200$}
	\psfrag{-100}[c][c]{\footnotesize -$100$}
	\psfrag{0}[c][c]{\footnotesize $0$}
	\psfrag{100}[c][c]{\footnotesize $100$}
	\psfrag{200}[c][c]{\footnotesize $200$}
	
	\psfrag{YYYYYYYYYYYYYYYY}{\footnotesize $\lvert H^{n^\mathrm{Tx}_\mathrm{A}}_{\mathrm{AFE}}(f)\rvert$ (dB)}
	\psfrag{XXXXXXXX}{\footnotesize $f$ (MHz)}
	
	\includegraphics[height=4.75cm]{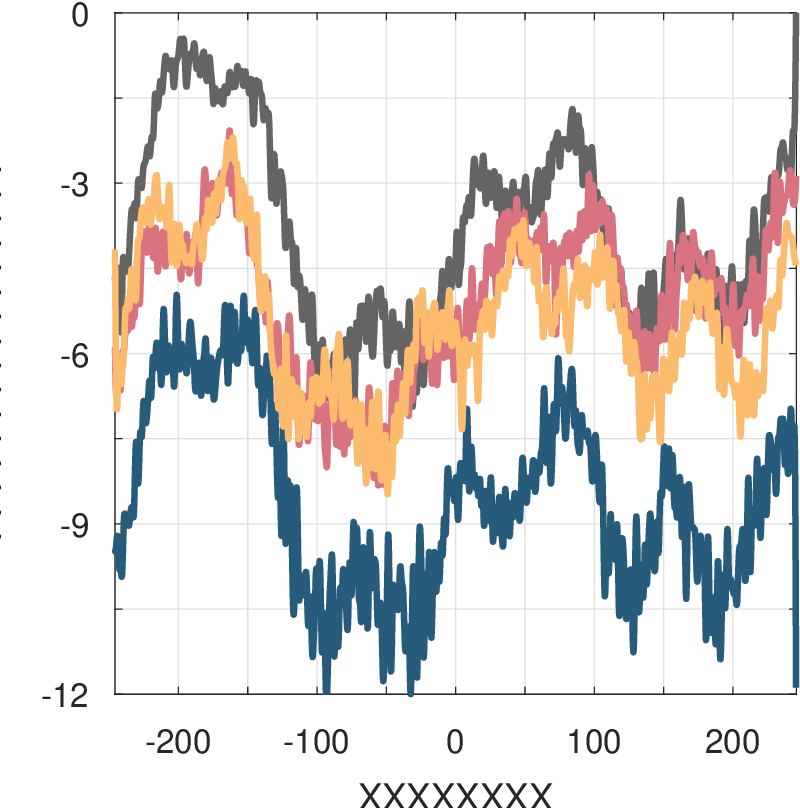}
	
	\captionsetup{justification=raggedright,labelsep=period,singlelinecheck=false}
	\caption{\ Normalized magnitude of the AFE CFR $H^{n^\mathrm{Tx}_\mathrm{A}}_{\mathrm{AFE}}(f)$ versus BB frequency $f$ for transmit channels \mbox{$n^\mathrm{Tx}_\mathrm{A}=0$} \mbox{({\color[rgb]{0.3922,0.3922,0.392}\rule[0.5ex]{1.25em}{1pt}})}, \mbox{$n^\mathrm{Tx}_\mathrm{A}=1$} \mbox{({\color[rgb]{0.1490,0.3569,0.4824}\rule[0.5ex]{1.25em}{1pt}})}, \mbox{$n^\mathrm{Tx}_\mathrm{A}=2$} \mbox{({\color[rgb]{0.8471,0.4510,0.4980}\rule[0.5ex]{1.25em}{1pt}})}, and \mbox{$n^\mathrm{Tx}_\mathrm{A}=3$} \mbox{({\color[rgb]{0.9882,0.7333,0.4275}\rule[0.5ex]{1.25em}{1pt}})}.}\label{fig:CFR_AFE}
	
\end{figure}
\begin{figure}[!t]
	\centering
	
	\subfloat[ ]{
		
		\psfrag{-12}[c][c]{\footnotesize -$12$}
		\psfrag{-9}[c][c]{\footnotesize -$9$}
		\psfrag{-6}[c][c]{\footnotesize -$6$}
		\psfrag{-3}[c][c]{\footnotesize -$3$}
		\psfrag{0}[c][c]{\footnotesize $0$}
		
		\psfrag{-200}[c][c]{\footnotesize -$200$}
		\psfrag{-100}[c][c]{\footnotesize -$100$}
		\psfrag{0}[c][c]{\footnotesize $0$}
		\psfrag{100}[c][c]{\footnotesize $100$}
		\psfrag{200}[c][c]{\footnotesize $200$}
		
		\psfrag{YYYYYYYYYYYYYYYY}{\footnotesize $\lvert H^{n^\mathrm{Rx}_\mathrm{A}}_{\mathrm{ABE}}(f)\rvert$ (dB)}
		\psfrag{XXXXXXXX}{\footnotesize $f$ (MHz)}
		
		\includegraphics[height=3.75cm]{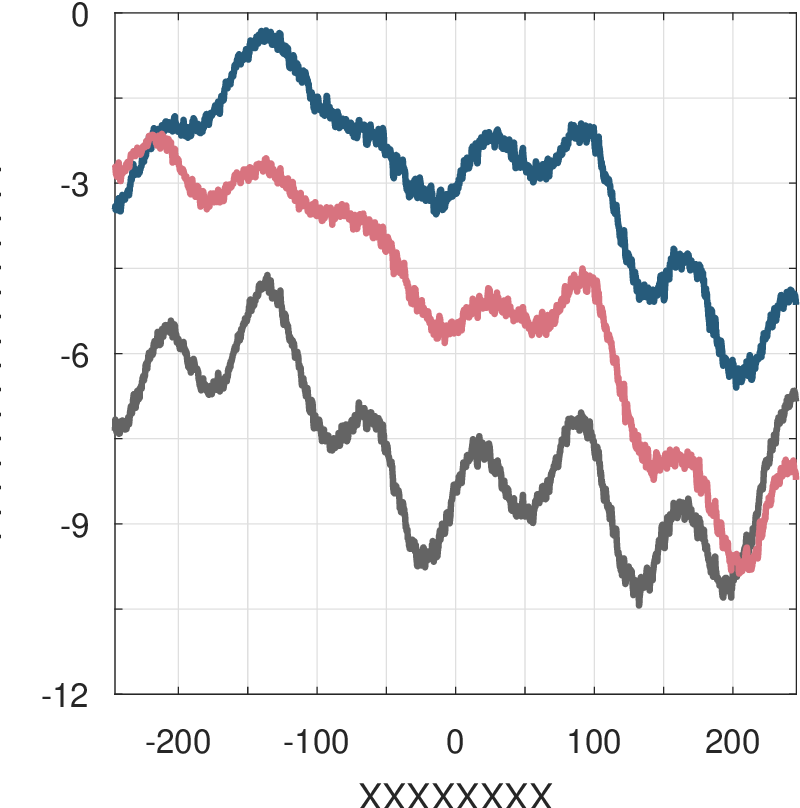}\label{fig:CFR_ABE}
		
	}\hspace{0.3cm}
	\subfloat[ ]{
		
		\psfrag{-20}[c][c]{\footnotesize -$20$}
		\psfrag{-15}[c][c]{\footnotesize -$15$}
		\psfrag{-10}[c][c]{\footnotesize -$10$}
		\psfrag{-5}[c][c]{\footnotesize -$5$}
		\psfrag{0}[c][c]{\footnotesize $0$}
		
		\psfrag{8}[c][c]{\footnotesize $8$}
		\psfrag{6}[c][c]{\footnotesize $6$}
		\psfrag{4}[c][c]{\footnotesize $4$}
		\psfrag{2}[c][c]{\footnotesize $2$}
		\psfrag{0}[c][c]{\footnotesize $0$}
		
		\psfrag{YYYYYYYYYYYYYYYY}{\footnotesize $\lvert h^{n^\mathrm{Rx}_\mathrm{A}}_{\mathrm{ABE}}(t)\rvert$ (dB)}
		\psfrag{XXXXXX}{\footnotesize $t$ (ns)}
		
		\includegraphics[height=3.75cm]{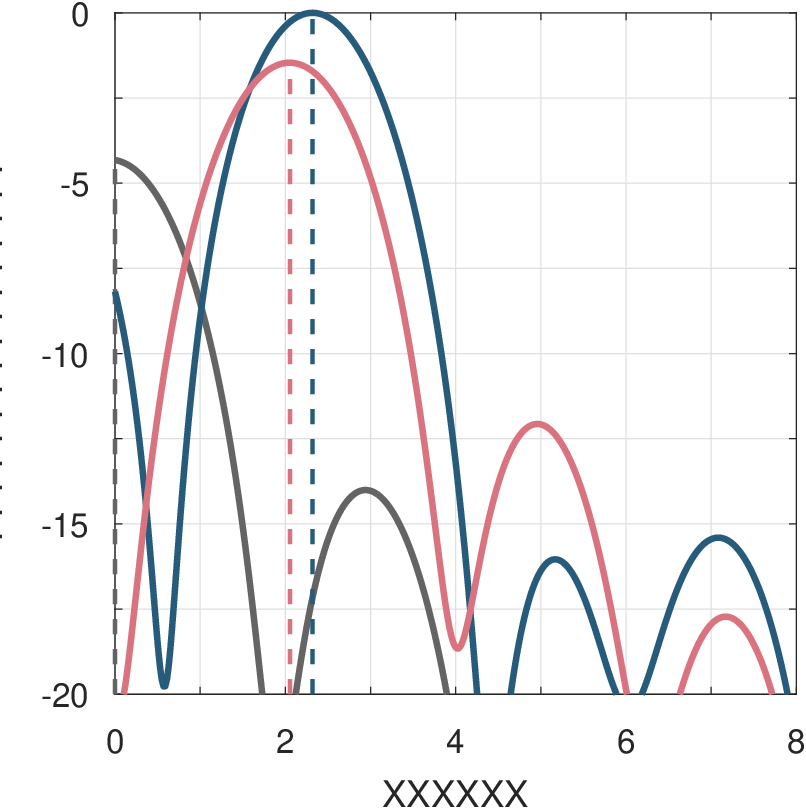}\label{fig:CIR_ABE}	
		
	}		
	
	\captionsetup{justification=raggedright,labelsep=period,singlelinecheck=false}
	\caption{\ Normalized magnitude of the ABE CFR $H^{n^\mathrm{Rx}_\mathrm{A}}_{\mathrm{ABE}}(f)$ versus BB frequency $f$ (a) and normalized magnitude of the ABE CIR $h^{n^\mathrm{Rx}_\mathrm{A}}_{\mathrm{ABE}}(t)$ versus time $t$ (b) for receive channels \mbox{$n^\mathrm{Rx}_\mathrm{A}=0$} \mbox{({\color[rgb]{0.3922,0.3922,0.392}\rule[0.5ex]{1.25em}{1pt}})}, \mbox{$n^\mathrm{Rx}_\mathrm{A}=3$} \mbox{({\color[rgb]{0.1490,0.3569,0.4824}\rule[0.5ex]{1.25em}{1pt}})}, and \mbox{$n^\mathrm{Rx}_\mathrm{A}=7$} \mbox{({\color[rgb]{0.8471,0.4510,0.4980}\rule[0.5ex]{1.25em}{1pt}})}. Their delays $\tau_\mathrm{ABE}^{n^\mathrm{Rx}_\mathrm{A}}$ associated with the peaks of the three shown \acp{CIR} are \mbox{$\tau_\mathrm{ABE}^{0}\approx\SI{0}{\nano\second}$} \mbox{({\color[rgb]{0.3922,0.3922,0.392}\rule[0.5ex]{0.5em}{0.5pt}\hspace{2pt}\rule[0.5ex]{0.5em}{0.5pt}})}, \mbox{$\tau_\mathrm{ABE}^{3}\approx\SI{2.32}{\nano\second}$} \mbox{({\color[rgb]{0.1490,0.3569,0.4824}\rule[0.5ex]{0.5em}{0.5pt}\hspace{2pt}\rule[0.5ex]{0.5em}{0.5pt}})}, and \mbox{$\tau_\mathrm{ABE}^{7}\approx\SI{2.06}{\nano\second}$} \mbox{({\color[rgb]{0.8471,0.4510,0.4980}\rule[0.5ex]{0.5em}{0.5pt}\hspace{2pt}\rule[0.5ex]{0.5em}{0.5pt}})}.}\label{fig:CFR_CIR_ABE}
	
\end{figure}

Before conducting the actual bistatic \ac{ISAC} measurements, the \acp{CFR} of the \mbox{$N^\mathrm{Tx}_\mathrm{A}=4$} transmit channels were estimated via calibration measurements with a reference receiver. Fig.~\ref{fig:CFR_AFE} shows the normalized magnitude of the \ac{AFE} \acp{CFR} $H^{n^\mathrm{Rx}_\mathrm{A}}_{\mathrm{ABE}}(f)$ as a function of the \ac{BB} frequency for all transmit channels. Their discrete-frequency domain equivalents were smoothed with a moving average filter and used to perform \ac{DPD}, ensuring beamforming towards the previously mentioned \acp{DoD}. In addition, the \acp{CFR} and the corresponding \acp{CIR} of all \mbox{$N^\mathrm{Rx}_\mathrm{A}=8$} receive channels of the \ac{ABE} were estimated via calibration measurements with a reference transmitter. Fig.~\ref{fig:CFR_CIR_ABE} shows the normalized magnitude of the \ac{ABE} \acp{CFR} $H^{n^\mathrm{Rx}_\mathrm{A}}_{\mathrm{ABE}}(f)$ as a function of the \ac{BB} frequency only for the receive channels \mbox{$n^\mathrm{Rx}_\mathrm{A}\in\{0,3,7\}$} for better visualization. Their equivalent discrete-frequency domain representations $H^{n^\mathrm{Rx}_\mathrm{A}}_{\mathrm{ABE},n,m}$ are later used for bistatic radar signal processing as described by \eqref{eq:D_radar}. In addition, the normalized magnitude of the corresponding \ac{ABE} \acp{CIR} $h^{n^\mathrm{Rx}_\mathrm{A}}_{\mathrm{ABE}}(t)$ are also shown. The delays of their dominant paths, i.e., $\tau_\mathrm{ABE}^{n^\mathrm{Rx}_\mathrm{A}}$, ultimately influence both the local \ac{STO} estimates at each channel as described in \eqref{eq:STO_channel} as well as the global \ac{STO} estimate calculated as in \eqref{eq:STO_est2}.

\begin{table}[!t]
	\renewcommand{\arraystretch}{1.5}
	\arrayrulecolor[HTML]{708090}
	\setlength{\arrayrulewidth}{.1mm}
	\setlength{\tabcolsep}{4pt}
	
	\centering
	\captionsetup{width=43pc,justification=centering,labelsep=newline}
	\caption{\textsc{Local and Global Synchronization Offset Estimates}}
	\label{tab:syncResults}
	\begin{tabular}{|c|c|c|c|}
		\hhline{|====|}
		\textbf{Receive channel}             & \textbf{Residual STO}                 & \textbf{CFO}                & \textbf{SFO}        \\ \hhline{|====|}
		$n^\mathrm{Rx}_\mathrm{A}=0$ & $\SI{0.0235}{\nano\second}$ & $\SI{15.4077}{\kilo\hertz}$ & $\SI{-4.1615}{ppm}$ \\ \hline
		$n^\mathrm{Rx}_\mathrm{A}=1$ & $\SI{0.4069}{\nano\second}$ & $\SI{15.3095}{\kilo\hertz}$ & $\SI{-4.1604}{ppm}$ \\ \hline
		$n^\mathrm{Rx}_\mathrm{A}=2$ & $\SI{1.4242}{\nano\second}$ & $\SI{15.3760}{\kilo\hertz}$ & $\SI{-4.1606}{ppm}$ \\ \hline
		$n^\mathrm{Rx}_\mathrm{A}=3$ & $\SI{1.6276}{\nano\second}$ & $\SI{15.5673}{\kilo\hertz}$ & $\SI{-4.1613}{ppm}$ \\ \hline
		$n^\mathrm{Rx}_\mathrm{A}=4$ & $\SI{1.0173}{\nano\second}$ & $\SI{15.6186}{\kilo\hertz}$ & $\SI{-4.1609}{ppm}$ \\ \hline
		$n^\mathrm{Rx}_\mathrm{A}=5$ & $\SI{1.2207}{\nano\second}$ & $\SI{15.4008}{\kilo\hertz}$ & $\SI{-4.1609}{ppm}$ \\ \hline
		$n^\mathrm{Rx}_\mathrm{A}=6$ & $\SI{1.4242}{\nano\second}$ & $\SI{15.5489}{\kilo\hertz}$ & $\SI{-4.1586}{ppm}$ \\ \hline
		$n^\mathrm{Rx}_\mathrm{A}=7$ & $\SI{2.0345}{\nano\second}$       & $\SI{15.5884}{\kilo\hertz}$ & $\SI{-4.1607}{ppm}$ \\ \hhline{|====|}
		\textbf{Global estimate}     & \textbf{N/A}                 & \textbf{$\SI{15.4772}{\kilo\hertz}$} & \textbf{$\SI{-4.1606}{ppm}$} \\ \hhline{|====|}
	\end{tabular}
\end{table}

To enable both communication and radar sensing processing, \ac{STO}, \ac{CFO} and \ac{SFO} were individually performed at each receive channel as described in Section~\ref{subsubsec:sync}. The obtained local synchronization offset estimates, as well as the resulting global estimates are shown in Table~\ref{tab:syncResults}. Since the earliest estimated \ac{OFDM} frame start point among all $N^\mathrm{Rx}_\mathrm{A}$ receive channels, which in the measurements was the one for receive channels \mbox{$n^\mathrm{Rx}_\mathrm{A}\in\{0,1\}$}, is taken as a common start point for all channels, only the residual \ac{STO} later estimated via pilots is shown. The obtained residual \ac{STO} values show that larger compensation was needed for the remaining receive channels, i.e., \mbox{$n^\mathrm{Rx}_\mathrm{A}=2$} through \mbox{$n^\mathrm{Rx}_\mathrm{A}=7$} as expected. Overall, a mean value of $\SI{1.17}{\nano\second}$ and standard deviation of $\SI{0.11}{\nano\second}$ was observed among the local residual \ac{STO} estimates, which respectively correspond to $57.50\%$ and $5.38\%$ of the critical sampling period \mbox{$T_\mathrm{s}=1/B$}. The \ac{CFO} estimates, in turn, include both the first estimation via the \ac{SC} algorithm and the fine tuning with pilots. A standard deviation of $\SI{116.18}{\hertz}$ was observed among the \ac{CFO} estimates, which corresponds to only $0.05\%$ of the subcarrier spacing of \mbox{$\Delta f=\SI{240}{\kilo\hertz}$}. This shows that the local estimates were already accurate and therefore converged to an even more acurate global \ac{CFO} estimate. As for \ac{SFO}, a standard deviation of only $\SI[parse-numbers=false]{48.45\times10^{-3}}{ppm}$ was observed among the local estimates, which was the case due to the use of the \ac{TITO} algorithm that allows performing highly accurate \ac{SFO} estimation \cite{giroto2024_tmtt}.

\begin{figure}[!t]
	\centering
	
	\subfloat[ ]{
		
		\psfrag{-1}[c][c]{\footnotesize -$1$}
		\psfrag{0}[c][c]{\footnotesize $0$}
		\psfrag{1}[c][c]{\footnotesize $1$}
		
		\psfrag{-2}[c][c]{\footnotesize -$1$}
		\psfrag{3}[c][c]{\footnotesize $0$}
		\psfrag{4}[c][c]{\footnotesize $1$}
		
		\psfrag{I}{\small $I$}
		\psfrag{Q}{\small $Q$}
		
		\includegraphics[height=3.75cm]{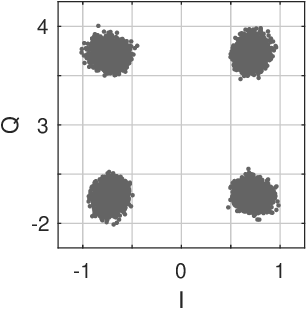}\label{fig:meas_const_a}
		
	}\hspace{0.1cm}
	\subfloat[ ]{
		
		\psfrag{-1}[c][c]{\footnotesize -$1$}
		\psfrag{0}[c][c]{\footnotesize $0$}
		\psfrag{1}[c][c]{\footnotesize $1$}
		
		\psfrag{-2}[c][c]{}
		\psfrag{3}[c][c]{}
		\psfrag{4}[c][c]{}
		
		\psfrag{I}{\small $I$}
		\psfrag{Q}{}
		
		\includegraphics[height=3.75cm]{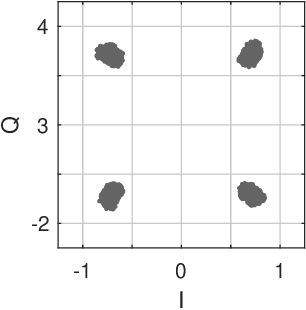}\label{fig:meas_const_b}
		
	}
	
	\captionsetup{justification=raggedright,labelsep=period,singlelinecheck=false}
	\caption{\ Measured QPSK constellations: (a) single channel with ZF equalization and (b) MRC of all $N^\mathrm{Rx}_\mathrm{A}=8$ receive channels.}\label{fig:meas_const}
\end{figure}
\begin{figure}[!t]
	\centering
	
	\subfloat[ ]{
		
		\psfrag{0}[c][c]{\footnotesize $0$}
		\psfrag{2}[c][c]{\footnotesize $2$}
		\psfrag{4}[c][c]{\footnotesize $4$}
		\psfrag{6}[c][c]{\footnotesize $6$}	
		
		\psfrag{-10}[c][c]{\footnotesize -$10$}
		\psfrag{-5}[c][c]{\footnotesize -$5$}
		\psfrag{0}[c][c]{\footnotesize $0$}
		\psfrag{5}[c][c]{\footnotesize $5$}
		\psfrag{10}[c][c]{\footnotesize $10$}
		
		\psfrag{0}[c][c]{\footnotesize $0$}
		\psfrag{-15}[c][c]{\footnotesize -$15$}
		\psfrag{-30}[c][c]{\footnotesize -$30$}
		\psfrag{-45}[c][c]{\footnotesize -$45$}
		\psfrag{-60}[c][c]{\footnotesize -$60$}
		
		\psfrag{Doppler shift (kHz)}[c][c]{\footnotesize Doppler shift (kHz)}
		\psfrag{Rel. bist. range (m)}[c][c]{\footnotesize Rel. bist. range (m)}
		\psfrag{Norm. mag. (dB)}[c][c]{\footnotesize Norm. mag. (dB)}
		
		\includegraphics[height=3.75cm]{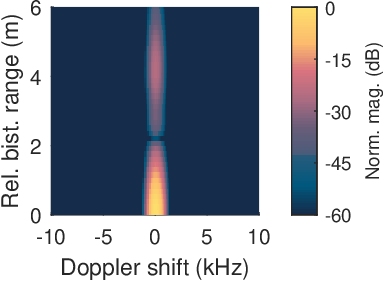}\label{fig:I_rD_meas}
		
	}\\
	\subfloat[ ]{
		
		\psfrag{0}[c][c]{\footnotesize $0$}
		\psfrag{2}[c][c]{\footnotesize $2$}
		\psfrag{4}[c][c]{\footnotesize $4$}
		\psfrag{6}[c][c]{\footnotesize $6$}	
		
		\psfrag{-40}[c][c]{\footnotesize -$40$}
		\psfrag{-20}[c][c]{\footnotesize -$20$}
		\psfrag{0}[c][c]{\footnotesize $0$}
		\psfrag{20}[c][c]{\footnotesize $20$}
		\psfrag{40}[c][c]{\footnotesize $40$}
		
		\psfrag{0}[c][c]{\footnotesize $0$}
		\psfrag{-10}[c][c]{\footnotesize -$10$}
		\psfrag{-20}[c][c]{\footnotesize -$20$}
		\psfrag{-30}[c][c]{\footnotesize -$30$}
		
		\psfrag{Azimuth (deg)}[c][c]{\footnotesize Azimuth (deg)}
		\psfrag{Rel. bist. range (m)}[c][c]{\footnotesize Rel. bist. range (m)}
		\psfrag{Norm. mag. (dB)}[c][c]{\footnotesize Norm. mag. (dB)}
		
		\includegraphics[height=3.75cm]{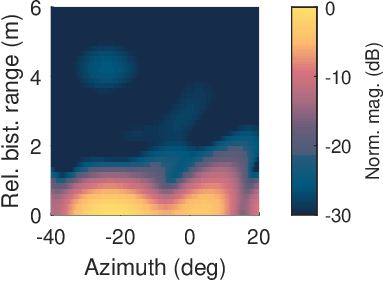}\label{fig:I_rA_meas}
		
	}
	
	\captionsetup{justification=raggedright,labelsep=period,singlelinecheck=false}
	\caption{\ Measured bistatic radar images: (a) range-Doppler shift radar image showing unresolved reflector and LoS paths at $\SI{0}{\meter}$ relative bistatic range, and (b) range-azimuth radar image for null Doppler shift showing resolved reflector and LoS paths at \ang{-22.5} and \ang{3}, respectively.}\label{fig:I_rDA_meas}
\end{figure}

Afterwards channel estimation and equalization were performed and the constellations shown in Fig.~\ref{fig:meas_const} were obtained. Specifically, Fig.~\ref{fig:meas_const_a} shows the receive \ac{QPSK} constellation for the channel \mbox{$n^\mathrm{Rx}_\mathrm{A}=0$} after \ac{ZF} \ac{FDE}, which has \ac{EVM} with a mean value of $\SI{-28.05}{dB}$ and $\SI{5.73}{dB}$ standard deviation. After \ac{MRC} following \eqref{eq:MRC} of all \mbox{$N^\mathrm{Rx}_\mathrm{A}=8$} receive channels, the constellation shown in Fig.~\ref{fig:meas_const_b} was obtained. This constellation is associated with a mean \ac{EVM} of $\SI{-28.98}{dB}$ and $\SI{5.34}{dB}$ standard deviation. After performing demodulation on the constellation from Fig.~\ref{fig:meas_const_b} and \ac{LDPC} decoding, followed by re-encoding and re-modulation, an estimate of the transmit \ac{OFDM} frame $\mathbf{X}$ was obtained and used for bistatic radar signal processing as described in \eqref{eq:D_radar}. First, range-Doppler radar images $\mathbf{I}^{n^\mathrm{Rx}_\mathrm{A}}$ were obtained for all receive channels, with an example for the channel \mbox{$n^\mathrm{Rx}_\mathrm{A}=0$} shown in Fig.~\ref{fig:I_rD_meas}. In this image, it can be seen that a single reflection is seen at a relative bistatic range of $\SI{0}{\meter}$ w.r.t. the \ac{LoS} reference path. This happens since the \ac{LoS} path and the reflector are unresolved, which is due to the fact that the difference of $\SI{15.23}{\centi\meter}$ between their bistatic ranges is smaller than the range resolution \mbox{$\Delta R=\SI{0.61}{\meter}$}. The second reflection seen at a relative bistatic range of $\SI{4.12}{\meter}$ is due to reflections in the room where the measurement was performed. After \ac{DoA} estimation via Fourier beamforming was performed as described in \eqref{eq:I_rDa}, a three-dimensional radar image $\bm{\mathcal{I}}$ was obtained. A cut of this image at a Doppler shift of $\SI{0}{\hertz}$ is shown in Fig.~\ref{fig:I_rA_meas}. In this image, it can be seen that the \ac{LoS} path and reflector are resolved, which confirms that the necessary phase coherence for \ac{DoA} estimation was achieved with the presented synchronization, communication and radar signal processing scheme for the proposed bistatic \ac{MIMO}-\ac{OFDM}-based \ac{ISAC} system in Section~\ref{sec:sys_model}. The deviations from the ground truth \acp{DoA} provided at the beginning of this section can be explained by limited accuracy and possible systematic errors in the ground truth measurements.


\section{Conclusion}\label{sec:conclusion}

This article introduced a system concept for bistatic \ac{MIMO}-\ac{OFDM}-based \ac{ISAC}. After a system model description, the additional processing steps w.r.t. a bistatic \ac{SISO}-\ac{OFDM}-based \ac{ISAC} system were described. These include transmit beamforming, a proposed distributed synchronization concept including \ac{STO}, \ac{CFO}, and \ac{SFO} estimation and correction, as well as bistatic radar signal processing to estimate range, Doppler shift, and \ac{DoA}.

In the proposed synchronization concept, it was shown that local offset estimates obtained at each receive channel can be averaged to obtain a single global estimate that is then used for synchronization at all channels in the \ac{CFO} and \ac{SFO} cases. As for \ac{STO}, different \acp{CIR} and consequently delays are experienced at each \ac{ABE} receive channel due to hardware non-idealities. Consequently, the earliest locally estimated \ac{OFDM} frame start point among all receive channels is set as the global start point estimate to avoid \ac{ISI}. The residual \acp{STO} can be later individually corrected at each receive channel based on estimates obtained via pilot subcarriers.

Focusing on \ac{STO} mismatch among the receive channels levels, simulation results showed that negligible communication performance degradation despite visibly increased \ac{EVM} as well as negligible peak and sidelobe distortion  in radar sensing were observed for an \ac{STO} standard deviation of $1\%$ the critical sampling period among the receive channels. At this point, however, a image \ac{SIR} drop of $\SI{10}{dB}$ compared to the case without \ac{STO} mismatch among the receive channels is experienced, which although moderate is not negligible. This level of imbalance among the \acp{STO} can, however, be corrected during \ac{STO} fine tuning based on pilots as sub-sample accuracy can be achieved. Finally, the claims and simulation results were confirmed by verification measurements with a $4\times8$ \ac{MIMO} setup at $\SI{27.5}{\giga\hertz}$. The results confirmed the similar \ac{CFO} and \ac{SFO} experienced by each individual receive channel, and also showed the mismatch in the \acp{STO}. All the aforementioned offsets could, however be estimated and corrected with the proposed synchronization approach, and communication and radar sensing capabilities in the proposed bistatic \ac{MIMO}-\ac{OFDM}-based \ac{ISAC} system concept were successfully demonstrated.


\bibliographystyle{IEEEtran}
\bibliography{./References/OverviewPapers,./References/RadCom_Enablement,./References/B5G_6G,./References/Interference,./References/Automotive,./References/RadarNetworks,./References/ChirpSequence,./References/PMCW,./References/OFDM,./References/OCDM,./References/OFDM_Variations,./References/CS_OCDM_Variations,./References/CP_DSSS,./References/BandwidthEnlargement_DigitalRadars,./References/CompressedSensing_DigitalRadars,./References/RadarTargetSimulator,./References/Parameters,./References/HardwareImplementation,./References/FirstRadCom,./References/Interference_CS,./References/ResourceAllocation,./References/SFO,./References/Bistatic,./References/RIS,./References/Nokia,./References/5G_NR,./References/CoMP,./References/LDPC,./References/PhaseNoise,./References/MIMOcomm,./References/MIMOradar}

\end{document}